\documentclass[final,authoryear,3p,times]{elsarticle}

\usepackage{graphicx}

\usepackage{amsmath,amssymb}

\newcommand{\bs}[1]{\boldsymbol{#1}} %bold symbol
\newcommand{\de}{\; \mathrm{d}}
\newcommand{\rmd}{\mathrm{d}}
\newcommand{\Tr}{\mathrm{tr}}
\newcommand{\trans}{\mathrm{T}}
\newcommand{\A}{A}
\newcommand{\E}{E}
\newcommand{\I}{I}

\usepackage[all]{xy}
\usepackage{longtable}

\newcounter{codecounter}[section]
\newcounter{eqcode}[codecounter]
\newenvironment{algenv}[2]  %% the updated algorithm environment
{
	\refstepcounter{codecounter}
	\noindent\rule{0.9\columnwidth}{1pt}\hfill\\[2ex]
	\parbox{.9\columnwidth}{
		{
			\bfseries Algorithm \thecodecounter:\hspace*{1ex}
		}
		\mbox{\textsc{#1}} \\
		{#2}
	}\\
	\noindent\rule{0.9\columnwidth}{1pt}\hfill%\\[2ex]
	\begin{longtable}
	{
		p{0.15\columnwidth}
		p{0.375\columnwidth}
		p{0.375\columnwidth}
	}
}
{
	\end{longtable}
	\vspace{-5ex}
	\noindent\rule{0.9\columnwidth}{1pt}\hfill\\
}

\newcommand{\twocolpf}[2]{
\multicolumn{2}{p{0.7\columnwidth}}{
#1
\vspace{-1ex}
\refstepcounter{eqcode}
\[
#2 \tag{\theeqcode}
\]
\vspace{-3ex}
}
}

\newcommand{\twocolp}[1]{
	\multicolumn{2}{p{0.75\columnwidth}}{
	#1
	}
}

\usepackage[usenames, dvipsnames]{xcolor}
\usepackage[breaklinks]{hyperref}

\journal{\hspace{-3cm}\colorbox{white}{\strut\hspace{4cm}\strut}}

\begin{document}

\begin{frontmatter}


\title{Sequential Monte Carlo EM for multivariate probit models\tnoteref{supp}}

%\tnotetext[supp]{Demonstration code for the sequential Monte Carlo EM algorithm is available as supplementary material online.}
\tnotetext[supp]{Demonstration code for the sequential Monte Carlo EM algorithm is available online at {\color{Cerulean!90!Blue} \url{https://github.com/annlia/smcemmpmDemo}}}

\author[gm]{Giusi Moffa}

\address[gm]{Institut f\"ur funktionelle Genomik, Universit\"at Regensburg, \\
Josef Engertstra\ss e 9, 93053 Regensburg, Germany}

\ead{giusi.moffa@ukr.de}

\author[jk]{Jack Kuipers}

\address[jk]{Institut f\"ur theoretische Physik, Universit\"at Regensburg, \\
D-93040 Regensburg, Germany}

\ead{jack.kuipers@ur.de}

\begin{abstract}
Multivariate probit models have the appealing feature of capturing some of the dependence structure between the components of multidimensional binary responses. The key for the dependence modelling is the covariance matrix of an underlying latent multivariate Gaussian. Most approaches to maximum likelihood estimation in multivariate probit regression rely on Monte Carlo EM algorithms to avoid computationally intensive evaluations of multivariate normal orthant probabilities. As an alternative to the much used Gibbs sampler a new sequential Monte Carlo (SMC) sampler for truncated multivariate normals is proposed. The algorithm proceeds in two stages where samples are first drawn from truncated multivariate Student $t$ distributions and then further evolved towards a Gaussian. The sampler is then embedded in a Monte Carlo EM algorithm. The sequential nature of SMC methods can be exploited to design a fully sequential version of the EM, where the samples are simply updated from one iteration to the next rather than resampled from scratch. Recycling the samples in this manner significantly reduces the computational cost. An alternative view of the standard conditional maximisation step provides the basis for an iterative procedure to fully perform the maximisation needed in the EM algorithm.  The identifiability of multivariate probit models is also thoroughly discussed.  In particular, the likelihood invariance can be embedded in the EM algorithm to ensure that constrained and unconstrained maximisation are equivalent. A simple iterative procedure is then derived for either maximisation which takes effectively no computational time. The method is validated by applying it to the widely analysed Six Cities dataset and on a higher dimensional simulated example. Previous approaches to the Six Cities dataset overly restrict the parameter space but, by considering the correct invariance, the maximum likelihood is quite naturally improved when treating the full unrestricted model. 
\end{abstract}


\end{frontmatter}

% \linenumbers

\section{Introduction}
\label{intro}
Multivariate probit models, originally introduced by \citet{art:AshfordS70} for the bivariate
case, are particularly useful tools to capture some of the dependence
structure of binary, and more generally multinomial, response variables
\citep{art:McCulloch94, art:McCullochR94, art:BockG96, art:ChibG98,
art:NatarajanMcCK2000, art:GueorguievaA2001, art:LiS2008}. 
Inference for such models is typically computationally involved and 
often still impracticable in high dimensions. 
To mitigate these difficulties, \citet{art:VarinC2010} proposed a 
pseudo-likelihood approach as a surrogate for a full likelihood analysis. 
Similar pairwise likelihood approaches were also previously considered 
by \citet{art:KukN2000} and \citet{art:RenardMG2004}. A number of Bayesian
approaches have also been considered including 
\cite{art:ChibG98, art:McCullochPR2000, art:Nobile98, art:Nobile2000, art:ImaiD2005} 
and more recently \cite{art:TalhoukDM2012}.

Due to the data augmentation or latent variable nature of the problem, the expectation maximisation
(EM) algorithm \citep{art:DempsterLR77} is typically employed for maximising the 
likelihood as its iterative procedure is usually more attractive than classical 
numerical optimisation schemes.  Each iteration consists of an expectation (E) step and a
maximisation (M) step, and both should ideally be easy to implement.  

For cases in which the E step is analytically intractable, \cite{art:WeiT90} introduced 
a Monte Carlo version of the EM algorithm (MCEM).  Sampling from the truncated normal 
distributions involved is often based on Markov chain Monte Carlo (MCMC) methods 
and the Gibbs sampler in particular \citep[see e.g.][]{art:Geweke91}.  
As a different option we employ a sequential Monte Carlo (SMC) sampler \citep{art:DelMoralDJ2006}. 
A sequence of distributions of interest is then approximated by a collection of
weighted random samples, called particles, which are progressively updated 
by means of sampling and weighting operations.
Though originally introduced in dynamical scenarios 
\citep{art:GordonSS93, art:Kitagawa96, art:LiuC98,SMCpractice} 
as a more general alternative to the well known Kalman filter \citep{art:Kalman60}, 
SMC algorithms can also be used in static inference \citep[see e.g.][]{art:Chopin2002} 
where artificial dynamics are introduced. 
When the target is a truncated multivariate normal, 
as in our case, an obvious sequence of distributions 
is obtained by gradually shifting the truncation region to the desired position. 
Since normal distributions decay very quickly in the tails, we propose to use 
flatter Student $t$ distributions to drive the SMC particles more efficiently towards 
the target region, and only then take the appropriate limit to recover the required 
truncated multivariate normal. The resulting algorithm is compared to the Gibbs 
sampler \citep{art:Geweke91, art:Robert95}.

The main difficulty in the M step rests with the computational complexity
of standard numerical optimisation over large parameter spaces,
for which \cite{art:MengR93} suggested a conditional maximisation approach. 
A simple extension of their method allows us to define an iterative procedure 
to further maximise the likelihood at each M step. Though the likelihood converges, 
there is no guarantee that the parameters converge to a point \citep{art:Wu83}. 
Restrictions to the parameter space have then been introduced to treat the identifiability 
issue where the data does not determine the parameters uniquely \citep{art:McCullochR94,art:BockG96}, 
raising the problem of constrained maximisation, normally significantly more difficult than unconstrained.
When constraints are only introduced to overcome identifiability issues rather than being intrinsic to
the problem, they can be regarded as artificial, and similar observations are at the basis
of parameter expansion approaches to EM \citep{art:LiuRW98}. 
In fact we show in our analysis of multivariate probit models that both constrained 
and unconstrained maximisation can be made identical. 
Furthermore we describe a simple novel strategy which allows 
either maximisation to be easily computed.

Building on the fundamental ideas of the SMC methodology it is possible to define a sequential version
of the EM, where the particle approximation is simply evolved after the parameter update in the M step, 
rather than resampled from scratch, so reducing the computational burden of the otherwise expensive
E step. Finally we validate our methods by comparison with previous approaches \citep{art:ChibG98,art:Craig2008}, 
and on a simulated higher dimensional example.
\section{Background and notation}
\subsection{Sequential Monte Carlo samplers} \label{sec:smcsampler}
Sequential Monte Carlo samplers \citep{art:DelMoralDJ2006} are a class of 
iterative algorithms to produce weighted sample approximations from a sequence 
$\{ \pi_n \}$ of distributions of interest where the normalising constant $C_n$ need not 
be known, $\pi_n= \gamma_n/C_n$.  For a given probability distribution $\pi$, 
one obtains a collection of weighted samples $\{ W^{(k)}, \bs{Z}^{(k)} \}$, also referred to
as particle approximation of $\pi$, such that
\begin{equation}\nonumber
\E_{\pi} \left ( h(\bs{Z}) \right) \simeq \sum_{k=1}^M W^{(k)} h (\bs{Z}^{(k)}),
\end{equation}
where $M$ is the number of particles and $h$ a function of interest.
In a static scenario the main purpose is to obtain such an approximation 
from the last element of an artificially defined targeted sequence.

In order to control for the degeneracy of the sample, resampling 
\citep[see][for a review of resampling schemes]{art:DoucCM2005} 
is typically performed  when the effective sample size (\textsc{ESS}), 
as defined by \cite{art:KongLW94} and often approximated as \citep{art:DoucetGA2000}: 
\begin{equation}
\textsc{ESS}^{-1} = \sum_{k=1}^M (W_n^{(k)})^2,
\label{eq:ESSdef}
\end{equation}
falls below a given threshold \textsc{ESS}$^{\star}=sM$, with $0<s<1$ though 
often $s=1/2$ is chosen as a trade-off between efficiency and accuracy.
The move from the target $\pi_{n-1}$ to the next $\pi_n$ 
is achieved by means of a transition kernel $K_n$, so that 
$Z_n^{(k)}  \thicksim K_n(Z_{n-1}^{(k)}, \cdot)$, and updating 
the normalised weights
\begin{equation}
W_n^{(k)} \propto W_{n-1}^{(k)} \tilde{w}_n ^{(k)}, \qquad
\tilde{w}_n(\bs{Z}_{n-1}^{(k)},\bs{Z}_n^{(k)}) = \frac{\gamma_n(\bs{Z}_n^{(k)}) L_{n-1}(\bs{Z}_n^{(k)}, \bs{Z}_{n-1}^{(k)})}
{\gamma_{n-1}(\bs{Z}_{n-1}^{(k)}) K_n(\bs{Z}_{n-1}^{(k)}, \bs{Z}_n^{(k)})}, \qquad k = 1, \ldots, M. \nonumber
\end{equation}
The quantity $L_{n-1}$ in the formula for the incremental weights $\tilde{w}_n ^{(k)}$ 
is a backward kernel introduced by \citet{art:DelMoralDJ2006} to address computational 
issues and should be optimised with respect to the transition kernel $K_n$ in order 
to minimise the variance of the importance weights. In the same work the authors 
also discuss a number of choices for $K_n$ suggesting MCMC kernels with $\pi_n$ 
as an invariant distribution as a convenient choice in many applications. A good 
approximation for the optimal backward kernel is then \citep{art:DelMoralDJ2006}
\begin{equation} \nonumber
L_{n-1}(\bs{z}_n, \bs{z}_{n-1}) = \frac{\pi_n(\bs{z}_{n-1}) K_n(\bs{z}_{n-1}, \bs{z}_n) }{\pi_n(\bs{z}_n) } .
\end{equation}
Following standard practice we therefore adopt here in particular a random walk 
Metropolis Hastings kernel. The samples at a given iteration $n$ are obtained by 
moving each particle $\bs{Z}_{n-1}^{(k)}$ to a new location 
$\bs{Z}_n^{(k)} = \bs{Y}^k \thicksim \mathcal{N}( \bs{Z}_{n-1}^{(k)}, \bs{\Sigma}^{\rm MH}_n)$ 
with probability $\alpha^k = 1 \wedge \rho^k$ and leaving it unchanged otherwise, 
with $\rho^k = \pi_n(\bs{Y}^k) / \pi_n(\bs{Z}_{n-1}^{(k)} )$. The covariance matrix 
$\bs{\Sigma}^{\rm MH}_n = \kappa \widehat{\bs{\Sigma}}_{\pi}$ in the random walk proposal is a scaled 
version of an approximation $\widehat{\bs{\Sigma}}_{\pi}$ of the target covariance matrix. 
In practice we set $\bs{\Sigma}^{\rm MH}_n = \kappa \widehat{\bs{\Sigma}}_{\pi_{n-1}}$ since at iteration 
$n$, $\pi_{n-1}$ is the best approximation available for $\pi_{n}$. Unlike MCMC schemes 
however, in the case of SMC samplers no convergence conditions are required since any 
discrepancies arising from sampling from the wrong distribution are corrected 
by means of importance sampling reweighting 
\citep[see e.g.\ section 2.2.3 of][]{art:DelMoralDJ2007}. 

As extensively investigated in the MCMC literature (for example the original paper of 
\citealp{art:GilksRS98, art:HaarioST2001, art:AtchadeR2005}, 
or the review of \citealp{art:AndrieuT2008}) the scaling factor $\kappa$ can be 
adaptively tuned by monitoring the average empirical acceptance probability $\hat{\alpha}_n$ 
at iteration $n$. For some canonical target distributions it has been proved 
by \cite{art:RobertsGG97} that the asymptotically (with the dimension) 
optimal acceptance rate is $0.234$, while \cite{art:RobertsR98} found that it 
is $0.574$ for Metropolis adjusted Langevin algorithms \citep[see also][for a survey of results]{art:RobertsR2001}. 
There is however no gold standard on how to choose the desired acceptance 
rate in more realistic situations. Within the SMC framework, where one of the 
purposes of the MCMC move is to help maintain the sample diversity it seems 
sensible to fix slightly higher values than those found in theory. Especially 
high values should nevertheless be avoided when performing local moves as they 
would only be masking an eventual sample degeneracy, and lead to highly 
correlated samples \citep[see][for a discussion of related issues]{art:Chopin2002}.  

In the case of SMC samplers with a Metropolis Hastings transition kernel, the 
empirical acceptance rate can be evaluated as
\begin{equation}\nonumber 
\hat{\alpha}_n = \sum_{k=1}^M W_n^{(k)} ( 1 \wedge \pi_n(\bs{Y}_n^{(k)}) / \pi_n(\bs{Z}_{n-1}^{(k)}) ).
\end{equation}
Hence a stochastic approximation type algorithm can be implemented aiming to 
keep the above quantity equal (or close) to a prespecified value $\alpha^{\star }$ 
\citep[see e.g.\ section 4.2 of][]{art:AndrieuT2008} by setting
$\bs{\Sigma}^{\rm MH}_{n+1} = \kappa_n \widehat{\bs{\Sigma}}_{\pi_{n}}$
with scaling factor adapted as
\begin{equation}
\log(\kappa_{n+1}) = \log(\kappa_{n}) + \xi_n (\hat{\alpha}_n(\log(\kappa_{n})) - \alpha^{\star }) ,
\end{equation}
with $\xi_n$ a stepsize and where the logarithm ensures that the scaling factors 
are positive.

Adaptation of the transition kernel specifically within SMC has also recently been 
considered by \citet{art:JasraSDT2011} and \citet{art:FearnheadT2013}.
\subsection{Monte Carlo EM}
An EM algorithm \citep{art:DempsterLR77} is an iterative procedure for the computation 
of maximum likelihood or maximum a posteriori estimates in the context of 
incomplete data problems, where the likelihood is typically intractable.
The algorithm relies on the definition of an associated complete data problem for which
the object function of the maximisation is tractable and therefore more easily solved. 
Let $\bs{Y}$ be a random variable representing the observed data and $\bs{\psi}$ a vector of unknown
parameters. Alternating between an expectation or E step and a maximisation or M step the algorithm
provides us with a sequence $\{\bs{\psi}^m\}$ of parameter estimates
such that the observed data likelihood $\mathcal{L}(\bs{\psi} \mid \bs{Y})$ is non decreasing (namely
$\mathcal{L}(\bs{\psi}^{m+1} \mid \bs{Y}) \geq \mathcal{L}(\bs{\psi}^m \mid \bs{Y})$), 
and eventually converges to a local maximum.
Let $\bs{Z}$ be a random variable corresponding to the augmented data. 
Separating the observed data log-likelihood in terms of the complete $(\bs{Y,Z})$ 
and conditional missing data $\bs{Z} \mid \bs{Y}, \bs{\psi}$ distributions 
and by taking the expectation with respect to the latent variable 
$\bs{Z} \mid \bs{Y}, \bs{\psi}^m$ conditioned on the observed data $\bs{Y}$
and the parameter estimate $\bs{\psi}^m$ at iteration $m$, the log-likelihood can be written as
\begin{equation} \nonumber
l(\bs{\psi} \mid \bs{Y}) = \log(\mathrm{pr} \{\bs{Y} \mid \bs{\psi}\}) = Q(\bs{\psi}, \bs{\psi}^m) - H(\bs{\psi}, \bs{\psi}^m) ,
\end{equation}
with
\begin{equation} Q(\bs{\psi}, \bs{\psi}^m) = \E_{\bs{Z} \mid \bs{Y}, \bs{\psi}^m} \left[ \log(\mathrm{pr} \{\bs{Y,Z} \mid \bs{\psi}\})
\right], \qquad
H(\bs{\psi}, \bs{\psi}^m) = \E_{\bs{Z} \mid \bs{Y}, \bs{\psi}^m} \left[
\log(\mathrm{pr} \{\bs{Z} \mid \bs{Y}, \bs{\psi}\}) \right] . \nonumber
\end{equation}
Jensen's inequality implies that $H(\bs{\psi}, \bs{\psi}^m) \leq H(\bs{\psi}^m, \bs{\psi}^m)$, 
so that the likelihood is certainly not decreased at each step if
$Q(\bs{\psi}^{m+1}, \bs{\psi}^m) \geq Q(\bs{\psi}^m, \bs{\psi}^m)$. 
An iteration of the EM algorithm then comprises the following two steps
\begin{description}
\item[E step.] Evaluate $Q(\bs{\psi}, \bs{\psi}^m)$.
\item[M step.] Maximise $Q(\bs{\psi}, \bs{\psi}^m)$ with respect to $\bs{\psi}$.
\end{description}
When it is not possible to perform the E step analytically a standard solution is given by the
MCEM  \citep{art:WeiT90} where the expectation in the E step is replaced by a 
Monte Carlo estimate
\begin{equation}\label{eq:Qfun}
Q(\bs{\psi}, \bs{\psi}^m) 
= \E_{\bs{Z} \mid \bs{Y}, \bs{\psi}^m} 
	\left[ \log(\mathrm{pr} \{\bs{Y,Z} \mid \bs{\psi}\}) \right]
\simeq \frac{1}{M} \sum_{k=1}^M \log(\mathrm{pr} \{\bs{Y,Z_k} \mid \bs{\psi}\}),
\end{equation}
with the samples $\bs{Z_k}$ drawn from the conditional distribution of the augmented
data $\bs{Z} \mid \bs{Y}, \bs{\psi}^m$.

For situations where the maximisation in the M step is not feasible, \cite{art:DempsterLR77}
suggested settling for a value that simply increases $Q(\bs{\psi}, \bs{\psi}^m)$ at each iteration,
and they termed the resulting procedure a \emph{generalised} EM algorithm \citep[see also][section 1.5.5]{EMalgorithm}. 
When the M step cannot be performed analytically, to overcome the difficulties associated 
with numerical maximisation, \cite{art:MengR93} suggested replacing the maximisation 
over the full parameter space by a multi-step conditional maximisation over 
several subspaces in turn. Ideally we wish to set $\bs{\psi}^{m+1}$ to the value of $\bs{\psi}$ which maximises
$Q(\bs{\psi}, \bs{\psi}^m)$, as required by the actual EM. In Section~\ref{Mstep} we show, 
for the first time, how such a value can easily be found for the multivariate probit model.
\subsection{Multivariate probit model}\label{sec:mvtprobit}
Following the formulation in \cite{art:ChibG98}, denote by $\bs{y}^{j}$ a binary
vector corresponding to the $j$th observation of a response variable $\bs{Y}^j$
with $p$ components.  Let $\bs{x}_i^j$ be a size $k_i$ column vector containing
the covariates associated to the $i$th component $y_i^j$ of the $j$th observation
$\bs{Y}^j$. The first element of the vector of covariates $\bs{x}_i^j$ can be set to 1 
to account for an intercept. Define the $j$th matrix of covariates
\begin{equation}\nonumber
\bs{X}^j \triangleq \mbox{diag}((\bs{x}_1^j)^{\trans}, \ldots, (\bs{x}_p^j)^{\trans} ) ,
\end{equation}
as a $p \times k$ block diagonal matrix, with $k = \sum_{i=1}^p {k_i}$.
A multivariate probit model with parameters $\bs{\beta} \in \mathbb{R}^k$ and 
$\bs{\Sigma}$, a $p \times p$ covariance matrix, can be specified by setting
\begin{equation} 
\mathrm{pr} \{ \bs{Y}^j=\bs{y}^j \mid \bs{X}^j, \bs{\beta}, \bs{\Sigma} \}
= \int_{A_1^j}\cdots \int_{A_p^j} \phi_p(\bs{z}^j;\bs{X}^j \bs{\beta},\bs{\Sigma}) \de \bs{z}^j, 
\qquad A_i^j = \left \{ \begin{array}{lcr} (0, \infty ) & \mbox{if} & y_i^j = 1  \\
( -\infty, 0 ]& \mbox{if} & y_i^j = 0 
\end{array} \right. ,
\label{eq:probz}
\end{equation}
where $\phi_p$ is the density function of a multivariate normal random variable
with mean vector $\bs{\mu} = \bs{X}^j \bs{\beta}$ and covariance matrix $\bs{\Sigma}$. 
The vector of regression coefficient is 
$\bs{\beta} = (\bs{\beta}_1^{\trans}, \ldots, \bs{\beta}_p^{\trans})^{\trans} $, 
with each subvector $\bs{\beta}_i \in \mathbb{R}^{k_i}$ corresponding to 
the $i$th component of the response variable. Naturally the situation where 
it is assumed that the same number of covariates are observed for each component
of the response variable and the vectors $\bs{\beta}_i$ are also taken to be all identical 
can be treated as a special case. When considering particular settings however, care must be taken
in reconsidering the model identifiability, as discussed in Section~\ref{sec:identifymore}.

The probit model can also be understood in terms of a continuous latent variable construction, 
where the binary response $\bs{Y}$ is obtained by discretization of a multivariate Gaussian variable
\mbox{$\bs{Z} \thicksim \mathcal{N}(\bs{X} \bs{\beta}, \bs{\Sigma})$}. 
The observations are then thought of as obtained from an unobserved sample of 
multivariate Gaussian vectors $\{ \bs{z}^1, \ldots, \bs{z}^N \}$
as $y_i^j = \I_{z>0}( z_i^j )$, where specifically $\bs{Z}^j \thicksim  \mathcal{N}(\bs{X}^j \bs{\beta}, \bs{\Sigma})$
and $\I$ is the indicator function.

The covariance matrix $\bs{\Sigma}$ is a crucial parameter for the multivariate
probit model since it accounts, though indirectly, for some of the dependence 
structure among the components of the response variable. 
The identity matrix corresponds to the assumption of 
independence and the model reduces to a collection of independent 
one dimensional probit models, for which the regression coefficients $\bs{\beta}$ 
can be easily estimated, component by component, and used as starting point for 
more elaborate inference strategies.  An alternative to the identity for the initial 
covariance matrix can be obtained \citep{art:EmrichP91} by pairwise approximations, which are likely 
however to lead to non positive definite matrices. `Bending' techniques \citep{art:HayesH81, art:Montana2005} 
are then necessary to ensure the positivity of the eigenvalues.
\subsubsection{Monte Carlo E step}
For the multivariate probit model, letting $\bs{\psi} = (\bs{\Sigma},\bs{\beta})$ be the parameter vector
and $\bs{Z}^j \thicksim \mathcal{N}(\bs{X}^j \bs \beta, \bs \Sigma)$ the latent variables,
the complete data log-likelihood function is
\begin{equation}
\log(\mathrm{pr} \{\bs{Y,Z} \mid \bs{\psi}\}) = \sum_{j=1}^N {\log \left [ \I_{A^j}(\bs{z}^j)
\phi(\bs{z}^j; \bs{X}^j \bs{\beta, \Sigma}) \right ]} .
\label{eq:completelike}
\end{equation}
Using the cyclicity of the trace and ignoring some normalising constants, 
the corresponding $Q(\bs{\psi}, \bs{\psi}^m)$ function can be written as
\begin{equation}
2 Q(\bs{\psi}, \bs{\psi}^m) = -N\log \vert \bs \Sigma \vert  - N \Tr \left[ \bs \Sigma^{-1} \bs S \right], \qquad 
\bs S = \frac{1}{N} \sum_{j=1}^N \E_{\bs{Z}^j \mid \bs{Y}^j, \bs{\psi}^m}
\left \{ (\bs{Z}^j - \bs{X}^j \bs{\beta})(\bs{Z}^j - \bs{X}^j \bs{\beta} )^{\trans} \right \},
\label{eq:qem}
\end{equation}
and for completeness a detailed derivation is provided in \ref{app:QfunProbit}.
The expression in \eqref{eq:qem} can be transformed into the one provided in \citet{art:ChibG98}
by using the cyclicity of the trace as in \eqref{eq:quadforms}, 
but the form in \eqref{eq:qem} is convenient for the maximisation.
The second term of \eqref{eq:qem} is analytically intractable
since it involves expectations with respect to high dimensional truncated
multivariate Gaussian densities. In a MCEM approach \citep{art:WeiT90} 
the expectations can be approximated as 
\begin{equation}
\E_{\bs{Z}^j \mid \bs{Y}^j, \bs{\psi}^m} \left \{ (\bs{Z}^j - \bs{X}^j \bs{\beta}
)(\bs{Z}^j - \bs{X}^j \bs{\beta} )^{\trans} \right \} 
\simeq \sum_{k=1}^M { W^{j (k)} (\bs{Z}^{j (k)} - \bs{X}^j \bs{\beta} )(\bs{Z}^{j (k)} - \bs{X}^j \bs{\beta} )^{\trans}}  ,
\label{eq:expest}
\end{equation}
over a weighted sample $\{ W^{j (k)}, \bs{Z}^{j (k)}  \}_{k=1}^M$
from $\pi(\bs{z}^j \mid \bs{y}^j,\bs{\psi}^m) = \textsc{TMN}(A^j,  \bs{X}^j \bs{\beta, \Sigma})$, 
a multivariate normal distribution truncated to the domain $A^j$. The weights should be normalised
$\sum_{k=1}^M W^{j (k)} =1$ and the samples themselves may be approximate, such as provided by MCMC or 
importance sampling based algorithms. In our analysis we suggest to use particle approximations 
provided by the SMC samplers, which are detailed in Section~\ref{sec:smctmn} for the truncated 
multivariate normal distribution. The particle approximations so obtained can also be updated in a 
sequential manner from one EM iteration to the next, without the need to redraw the 
complete sample from scratch at each E step. The result is a more efficient EM algorithm 
as presented in Section~\ref{sec:smcem} for multivariate probit models.
\subsubsection{Two-step conditional maximisation}\label{sec:2stepcm}
The multivariate normal regression with incomplete data is considered as
an example in \cite{art:MengR93}. The parameters $\bs{\psi}^m$ at step $m$ 
are split into $\bs{\Sigma}^m$ and $\bs{\beta}^m$ leading to a two-step 
conditional maximisation which can be performed analytically.
The solutions can be obtained by setting to zero the derivatives of \eqref{eq:qem}. 
Using the cyclicity of the trace the maximisation condition becomes
\begin{equation} \label{eq:diffQ}
2\rmd Q = -N \Tr \left[\bs \Sigma^{-1} \left(\bs I-\bs S \bs \Sigma^{-1} \right)\rmd \bs \Sigma 
+ \bs \Sigma^{-1} \rmd \bs S \right]  = 0 ,
\end{equation}
The function $\bs{S}$ from \eqref{eq:qem} only depends on $\bs{\beta}$ so by fixing $\bs{\beta}$, 
the value of $\hat{\bs{\Sigma}}$ which satisfies equation \eqref{eq:diffQ} is simply 
$\bs{S}$.  Writing the result in terms of the particle approximation in \eqref{eq:expest}, we have
\begin{equation}
\hat{\bs{\Sigma}} \left(\bs{\beta}\right) = \frac{1}{N} \sum_{j=1}^N { \sum_{k=1}^M { W^{j (k)}  
(\bs{Z}^{j (k)} - \bs{X}^j \bs{\beta} )
(\bs{Z}^{j (k)} - \bs{X}^j \bs{\beta} )^{\trans} } } .
\label{eq:sigmaopt}
\end{equation}
For example, by evaluating this at the current regression vector value $\bs{\beta}^{m}$ 
the covariance matrix can be updated to $\bs{\Sigma}^{m+1}=\hat{\bs{\Sigma}}\left(\bs{\beta}^{m}\right)$ 
for the next step of the EM algorithm. Keeping $\bs{\Sigma}$ fixed, the cyclicity of 
the trace allows us to write the conditional maximisation condition from \eqref{eq:diffQ} as
\begin{equation} \label{eq:diffbeta}
0 = -N \Tr \left[\bs{\Sigma}^{-1} \rmd \bs{S} \right] \simeq 2 \left(\rmd \bs{\beta} \right)^{\trans}\sum_{j=1}^N \sum_{k=1}^M W^{j (k)}  
\left(\bs{X}^j \right)^{\trans} \bs{\Sigma}^{-1} (\bs{Z}^{j (k)} - \bs{X}^j \bs{\beta} ) ,
\end{equation}
where again the Monte Carlo estimate in \eqref{eq:expest} has been substituted for $\bs S$.
The value of $\hat{\bs{\beta}}$ which satisfies this condition is then
\begin{equation} \label{eq:betaopt}
\hat{\bs{\beta}} \left(\bs{\Sigma}\right)= 
\bigg ( \sum_{j=1}^N { (\bs{X}^j)^{\trans} \bs{\Sigma}^{-1}  \bs{X}^j }  \bigg )^{-1} 
\sum_{j=1}^N { (\bs{X}^j)^{\trans} \bs{\Sigma}^{-1} 
\sum \limits _{k=1}^M {  \left ( W^{j (k)} \bs{Z}^{j (k)} \right ) }  } ,
\end{equation}
so that by using the already updated value $\bs{\Sigma}^{m+1}$ the regression parameters 
for the next step can be updated as $\bs{\beta}^{m+1}=\hat{\bs{\beta}}\left(\bs{\Sigma}^{m+1}\right)$
to give the new parameters $\bs{\psi}^{m+1}$.  Though this two-step
approach does not maximise $\bs{\psi}$ at each step, it removes the need for
computationally intensive maximisation and (in the large $M$ limit with particle approximations) 
increases the likelihood at each step to ensure convergence of the (generalised) EM.
\subsection{Model invariance and identifiability}\label{sec:identify}
When the data is `incomplete', maximisation of the observed data likelihood may not
lead to uniquely identified parameters. Imposing constraints is a standard measure to 
ensure identifiability, but often with the effect of making the M step more involved 
\citep[e.g][and more specifically for multivariate probit models \citet{art:BockG96,art:ChanK97}]{art:KukC2001}. 
The phenomenon is directly linked to symmetries of the likelihood, where 
it is invariant under some change of coordinates of the parameters. 
Both \emph{global} and \emph{local} symmetries can play a role. 
In the first case the invariance of the likelihood $\mathcal{L}(\bs{\psi})$ 
does not depend on the particular value of $\bs{\psi} \in \Psi$. 
The parameter space can then be decomposed as $\Psi=\Delta\times\Xi$ 
into an invariant space $\Delta$ and a reduced parameter space 
$\Xi$ so that $\bs{\psi}=(\bs{\delta},\bs{\xi})$ with $\bs{\delta}\in\Delta$ and $\bs{\xi}\in\Xi$.  
Due to the invariance of the likelihood over $\Delta$
\begin{equation} \nonumber
\mathcal{L}(\bs{\psi})=\mathcal{L}(\bs{\delta},\bs{\xi})=\mathcal{L}(\bs{\xi}) \Rightarrow
\max_{\bs{\psi}}\mathcal{L}(\bs{\psi})=\max_{\bs{\xi}}\mathcal{L}(\bs{\xi}) ,
\end{equation}
unconstrained maximisation over the whole space $\Psi$ is identical to performing it
`constrained' over the reduced space $\Xi$, with the difference that the parameters 
maximising the likelihood in the larger space are $\bs{\psi}^*=\Delta\times\bs{\xi}^*$. 
Conversely, if the likelihood depended on some subspace of $\Delta$ then it would be 
identified during the maximisation process. Therefore the dimension of $\Delta$ is the 
number of constraints needed to ensure identifiability.

In addition to any global symmetries, the likelihood function could also show a 
\emph{local} symmetry so that $\hat{\mathcal{L}}(\bs{\xi})$ is maximised by a higher 
dimensional manifold rather than a single point \citep[as discussed in][]{art:Wu83}.  
In principle a local change of variables is possible (for example making the non-zero 
eigenvalues of the Hessian equal to $-1$ around the maximum) to decompose the space 
further, but in practice this presumes knowledge of the likelihood function.  As above
though, maximisation over the subspace or the whole space are exactly equivalent
because there will still be (local) dimensions which do not affect the value of the
likelihood.

Within the EM algorithm the identifiability issue becomes more subtle since the 
likelihood is not maximised directly, but by proxy through the function $Q(\bs{\psi},\bs{\psi}^m)$. 
If this were to share the symmetries of the likelihood, then the simpler unconstrained 
maximisation would be equivalent to the constrained version, as for the likelihood. 
If this is not the case, for example due to conditioning on the previous parameter 
value $\bs{\psi}^m$, then any changes in $Q$ arising from shifting $\bs{\psi}$ in the invariant 
space $\Delta$ of the likelihood must be exactly mimicked by changes in $H$.  This 
spurious dependence can create differences between constrained and unconstrained maximisation.
The non decreasing behaviour of the likelihood remains preserved, since neither maximisation 
decreases $Q$ nor, because of Jensens's inequality, increases $H$. Hence either choice leads 
to the EM algorithm finding a maximum of the likelihood (though not necessarily the same one) 
and explains the conjecture of \cite{art:BockG96,art:ChanK97} and the agreement between constrained and 
unconstrained maximisation found in \cite{art:KukC2001}.

In fact such symmetries can be seen as a natural example of the parameter expansion EM algorithm 
of \cite{art:LiuRW98} where in general one seeks additional parameters which do not affect the 
likelihood but which can be incorporated into the EM steps.  Here for example the standard 
EM would be over the constrained space $\Xi$ while the parameter expanded version would include some 
or all the parameters in $\Delta$.  By examining the 
second differentials of the likelihood, \cite{art:LiuRW98} showed that (at least in a 
quadratic neighbourhood of the maxima) the parameter expanded EM algorithm converges
at least as fast as the standard version, suggesting the more parameters the better.  
They also gave some examples where the speed up was very significant.  In general unconstrained 
maximisation is also less demanding than constrained maximisation and then more appealing 
on both fronts. 
\section{Methodology}
Most approaches to MCEM for multivariate probit \citep[e.g.][]{art:ChanK97, art:NatarajanMcCK2000} 
rely on MCMC schemes based on the Gibbs sampler to approximate the expectations in \eqref{eq:expest}. 
As an alternative in Section~\ref{sec:smctmn} we propose a SMC sampler for truncated multivariate normal distributions. 
In Section~\ref{sec:smcem} we discuss how to evolve the particle approximation through EM iterations and so
avoid to fully draw a new sample at each E step. By taking an alternative view of the conditional maximisation
an iterative procedure to complete the maximisation is discussed in Section~\ref{sec:mstepmp}.
Based on identifiability considerations specifically for multivariate probit models,
it is discussed in Section~\ref{sec:identifymore} how to perform constrained maximisation
at almost no computational cost.
\subsection{SMC sampler for truncated multivariate normal and $t$ distributions}\label{sec:smctmn}
Since the probability of a random walk Metropolis to move towards the tails of a
Gaussian distribution decreases exponentially,
a SMC method involving normals may be highly inefficient in moving samples 
towards regions of low probability.  To achieve higher rates of acceptance 
in the tails we suggest starting with a flatter distribution: the multivariate 
(of dimension $p$) Student $t$ distribution $\mathcal{T}(\nu, \bs{\mu}, \bs{\Sigma})$ 
with degree of freedom $\nu$, a size $p$ vector $\bs{\mu}$ and 
a $p\times p$ positive definite matrix $\bs{\Sigma}$ as location and scale parameters respectively. 
The probability density function of a variable $\bs{Z} \thicksim \mathcal{T}(\nu, \bs{\mu}, \bs{\Sigma})$
can be defined \citep{art:NadarajahK2005} as
\begin{equation}\label{eq:student}
f(\bs{z}) = \frac{\Gamma(\frac{\nu + p}{2})}{\Gamma(\frac{\nu}{2}) (\pi
\nu)^{p/2} \vert \bs{\Sigma} \vert^{1/2}} 
\left[ 1 + \frac{1}{\nu}(\bs{z}-\bs{\mu})^{\trans} \bs{\Sigma}^{-1}(\bs{z}-\bs{\mu}) \right]^{-\frac{\nu + p}{2}} .
\end{equation}
Replacing the $\nu$ in the denominator inside the square brackets by $(\nu-2)$, 
and correspondingly changing the normalisation factor, would provide the 
Student distribution with a covariance of $\bs{\Sigma}$.  As it stands, the 
distribution in \eqref{eq:student} actually has a covariance of 
$\nu\bs{\Sigma}/(\nu-2)$ which further increases the acceptance in the tails.
Once in the region of low probability we allow the number of degrees of
freedom to grow to infinity $(\nu \to \infty)$ so the distribution approaches a
$p$-variate Gaussian with the same mean and covariance matrix $\bs{\Sigma}$.  

To sample in the region of interest $A$, we define a sequence of target distributions 
$\{ \pi_n \}_0^T$ such that the first target is an unconstrained multivariate Student 
and the last one is the same distribution truncated to $A$. Quite naturally 
the intermediate distributions are defined in terms of intermediate target domains 
$\{ \A_n \} _0^T$, included in each other $\A_{k+1}\subset\A_{k}$, with $\A_T \equiv \A$ 
and $ \A_0 \equiv \mathbb{R}^p$. The local target $\pi_n$ at iteration 
$n$ of the SMC algorithm is then $\pi_n(\bs{z}) = {\gamma_n( \bs{z})}/{C_n}$, with
\begin{equation} \nonumber
\gamma_{n}(\bs{z}) =  \left[1 + \frac{1}{\nu}(\bs{z}-\bs{\mu})^{\trans}\bs{\Sigma}^{-1}(\bs{z}-\bs{\mu})
\right]^{-\frac{\nu + p}{2}}\I_{\A_n }(\bs{z}),
\end{equation}
where $C_n$ is a normalising constant which can be estimated \citep{art:DelMoralDJ2006} from
\begin{equation} \nonumber 
\widehat{C}_n = C_0 \prod_{i=1}^n \widehat{\frac{C_i}{C_{i-1}}} , \qquad
\widehat{\frac{C_i}{C_{i-1}}} = \sum_{k=1}^M W_{i-1}^{(k)}
\tilde{w}_i(\bs{Z}_{i-1}^{(k)}, \bs{Z}_i^{(k)}) ,
\end{equation}
and $C_0$ follows from \eqref{eq:student}. It follows that the probability 
that a random variable $\bs{Z} \thicksim \pi_0$ from the initial distribution 
falls within region $\A_n$ can be approximated by 
$P(\bs{Z} \in A_n) = \frac{C_n}{C_0} \simeq \prod_{i=1}^n \widehat{\frac{C_i}{C_{i-1}}}$. 
This ultimately allows us to obtain the probabilities of the regions in \eqref{eq:probz} 
and hence the likelihood for the probit model.

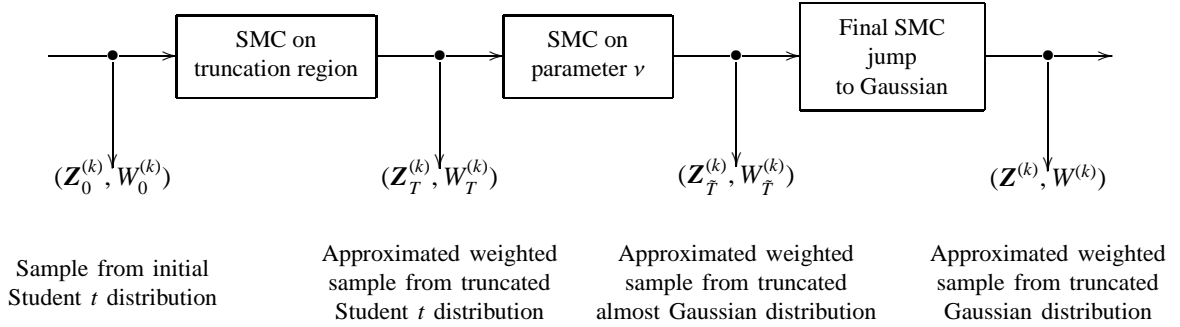
\begin{figure}
\begin{equation*}
\xymatrix@C=5pt{
\ar@{-}[r] &*{\bullet} \ar [d] \ar[r]
& *++[F] {\parbox{2.2cm}{\centering \small{SMC on \\ truncation region}}} \ar '[r] [rr] 
& *{\bullet} \ar [d]  
& *++[F] {\parbox{1.8cm}{\centering \small{SMC on \\ parameter $\nu$}}} \ar '[r] [rr] 
& *{\bullet} \ar [d] 
& *++[F] {\parbox{2cm}{\centering \small{Final SMC jump \\ to Gaussian}}} \ar '[r] [rr] 
& *{\bullet} \ar [d]
& {}
\\
& *-{ ( \bs{Z}_0^{(k)}, W_0^{(k)} )} && *-{ ( \bs{Z}_{T}^{(k)}, W_{T}^{(k)} )}  
&& *-{  ( \bs{Z}_{\tilde{T}}^{(k)}, W_{\tilde{T}}^{(k)} )} && *-{  ( \bs{Z}^{(k)}, W^{(k)} )}
\\
&\save[]+<0cm,-0.3cm>*-\txt<8pc>{%
\small{Sample from initial Student $t$ distribution}} \restore
&& \save[]+<0cm,-0.3cm>*-\txt<8pc>{%
\small{Approximated weighted sample from truncated Student $t$ distribution}} \restore
&&  \save[]+<0cm,-0.3cm>*-\txt<9pc>{%
\small{Approximated weighted sample from truncated almost Gaussian distribution}} \restore
&& \save[]+<0cm,-0.3cm>*-\txt<8pc>{%
\small{Approximated weighted sample from truncated Gaussian distribution}} \restore
}
\end{equation*}
\caption{Cascade interpretation of the SMC sampler for truncated multivariate normal via Student $t$.}
\label{fig:cascade}
\end{figure}

After reaching the required region, we define a new sequence
of target distributions $\{ \tilde{\pi}_n \}_0^{\tilde{T}}$ which this time 
starts from the truncated Student $\tilde{\pi}_0 = \pi_T = \gamma_T/C_T$. 
The following terms of the sequence are defined by increasing the degree 
of freedom $\nu$ up to a value $\nu_{\tilde{T}}$ large enough so that the 
truncated Student $\tilde{\pi}_{\tilde{T}} = \tilde{\pi}(\nu_{\tilde{T}})$ cannot be distinguished 
from the desired truncated multivariate normal within a certain level of 
accuracy. A final step is then performed to explicitly move to the Gaussian. 
One could also vary both the truncation region and the degree of freedom 
concurrently in the sequence of target distributions, but since the main reason 
for introducing the flatter Student distribution is to aid moving to regions 
of low probability we chose this two-step approach. 
A graphical overview of the process of moving from a Student $t$ distribution 
to a truncated Gaussian is given in Figure~\ref{fig:cascade}.

Similar sampling problems have also been studied in the literature dealing with rare event analysis. 
There smooth sequences of distributions gradually concentrating on the rare set have been suggested, 
as opposed to simply increasingly truncated distributions \citep[see][and references therein]{art:JohansenDMD2006}. 
The two stage approach based on the Student $t$ distribution is preferred here for its relative conceptual simplicity and the fact
that it proved well suited in practice to a linear adaptation framework of the type described in the following section \ref{sec:aaad}.

\subsubsection{Adaptive approach to artificial dynamics}\label{sec:aaad}
Adaptive strategies can be applied not only for tuning the transition kernel $K_n$, 
as noticed in Section~\ref{sec:smcsampler}, but also to define the artificial dynamics 
leading to the distribution of interest $\pi_T$. 
The problem of finding the optimal path linking an initial measure $\pi_0$ to the target $\pi_T$ 
on the space of distributions is not addressed, rather it is assumed that the functional form of the 
intermediate distributions is given and can be described in terms of a parameter $\theta$. 
In the examples of the Section~\ref{sec:smctmn} we have $\theta=A$ for the truncation case and 
$\theta=\nu$ when moving the truncated Student to a Gaussian. 
An adaptive strategy to move from $\pi_0$ to $\pi_T$ is one that does not require the sampling points 
$\{ \theta_n \}$ defining the intermediate targets $\{ \pi_n = \pi(\theta_n) \}$ to be fixed a priori, but 
allows us to determine them dynamically on the basis of the local difficulty of the problem.

Adaptation can be achieved by controlling some statistics related to the performance 
of the algorithm and evolving with the parameter $\theta$. The \textsc{ESS} introduced 
in \eqref{eq:ESSdef} is an ideal quantity to monitor. Theoretically we wish to solve
\begin{equation} \label{eq:ESS}
\textsc{ESS}_n(\theta_n) - \textsc{ESS}^{\star}_A = 0,
\end{equation}
where $\textsc{ESS}^{\star}_A$ is a value chosen to compromise between efficiency and accuracy, 
and which can be different (lower) from the resampling threshold \textsc{ESS}$^{\star}$. 
Inspired by the Robbins-Monro recursion \citep[see for example][page 3]{bk:KushnerY2003} 
for stochastic approximation, and aiming at the dynamical design of a sequence which keeps 
the \textsc{ESS} on average close to the threshold \textsc{ESS}$^{\star}_A$, we define the updating scheme
\begin{equation}
\theta_{n+1} = \left[
		\theta_{n} + 
			\left( 
				\zeta_n 
 					\frac{\widetilde{\mbox{ESS}}_n -
\mbox{ESS}^{\star}_A} {M}
				\vee \Delta \theta_{\min}
			\right) \right]
					\wedge \theta_T ,
\label{eq:saESS}
\end{equation}
where $\widetilde{\textsc{ESS}}_n$ is the value observed for \textsc{ESS}
at iteration $n$ and the division by the number of particles $M$ is only introduced for scaling 
purposes. Taking the maximum between the correction term and $\Delta \theta_{\min}$ ensures 
that the resulting sequence approaches the final target monotonically, while taking the 
minimum with $\theta_T$ ensures that the sequence ends at the desired target $\pi(\theta_T)$. 
Theoretically the \textsc{ESS} should ideally be equal to the total number of particles $M$ 
of the SMC sampler. To promote motion and so a quicker progression of the algorithm 
towards its final target, the threshold \textsc{ESS}$^{\star}_A$ 
can be fixed as a fraction $a \in (0,1)$ of $M$, namely \textsc{ESS}$^{\star}_A = aM$. 
The fraction $a$ should be slightly smaller than the fraction $s$ defined in 
\ref{sec:smcsampler} to control the resampling, 
say $a=.9s$, to ensure that the resampling 
threshold \textsc{ESS}$^{\star}$ is also crossed while the algorithm runs. 
The number of iterations needed to reach the target $\pi_T$ is reduced for smaller $a$.  
Similar adaptive ideas have also been applied to inference for 
stochastic volatility models by \citet{art:JasraSDT2011} and rather recently 
discussed in \citet{art:SchaeferC2013} for sampling in large binary spaces.

The details of the complete SMC sampler are summarised in Algorithm~\ref{alg:SMCsampler} in \ref{app:algorithms}.
\subsubsection{Scaling behaviour}
The advantage of the SMC method, over alternatives which may be more efficient
in sampling from truncated multivariate normals in low dimensions, is the scaling 
behaviour with the dimension $p$. Solving the adaptive equation \eqref{eq:ESS} 
exactly means that we lose a fixed proportion of the probability mass at each iteration.  
The number of steps required to reach a target region of low probability $r$, then
behaves like $\log(r)$, independently of $p$.  This may not be true when using
\eqref{eq:saESS} as a numerical adaptive approximation to \eqref{eq:ESS}, especially 
as the number of steps for the adaption to settle grows linearly with $p$, 
so a weak dependence on the dimension could be expected.

\begin{figure}
\begin{center}
  \includegraphics[width=0.6\textwidth]{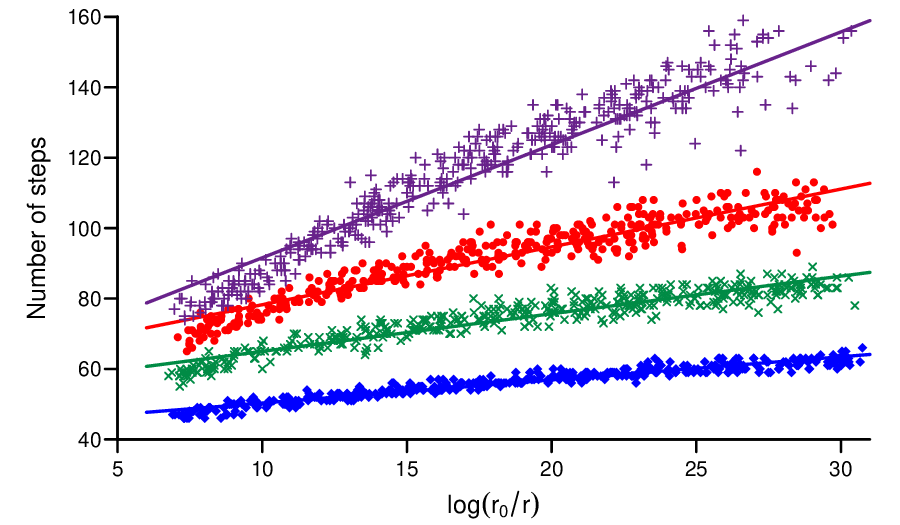}
\end{center}
\caption{The number of steps required for the SMC algorithm to reach a region of
probability $r$ for dimensions 2 (diamonds), 4 (crosses), 8 (dots) and 16 (pluses).}
\label{fig:scaling}
\end{figure}

A simulation study with targets of dimensions $2^n$ for $n=1,\ldots,4$ 
was performed. To limit the sources of variability, only 
one covariance structure was considered for the unconstrained distribution, with 
unit diagonals and a single non-zero off-diagonal element of 0$\cdot$9. The SMC 
algorithm was initialised so that after an initial move the Student $t$ 
target would be truncated to a region containing one quarter of the probability mass 
of an independent Gaussian, and we denote by $r_0$ the actual estimated probability. 
The cutoff for the final target, the same in all directions, was drawn so as to ensure 
that the log probability of an independent multivariate normal would be uniform on a given 
interval. The number of steps needed to reach the target are plotted against $\log(r_0/r)$
in Fig.~\ref{fig:scaling}, for 400 runs of a SMC sampler with 4000 particles for the 
different dimensions. A behaviour close to linear can be observed, though the offset 
increases by a factor of about 1$\cdot$4 over the range of dimensions and the slope 
increases roughly linearly with $p$, which is likely due to any inexactness in the 
adaptation.  The theoretical stability of these types of algorithms has recently 
been investigated in depth by \cite{art:BeskosCJ2012}.

\subsection{Sequential Monte Carlo EM for multivariate probit models}\label{sec:smcem}
The SMC sampler of Section~\ref{sec:smctmn} for truncated multivariate 
normals has good scaling behaviour for rare events, but depending on the choice of initial 
parameters $\bs{\psi}^0 = (\bs{\Sigma}^0,\bs{\beta}^0)$ there may be more efficient methods of 
obtaining the initial sample. If for example, as suggested earlier, the covariance 
matrix $\bs{\Sigma}^0$ is set to be the identity matrix $\bs{I}$, a sample of the corresponding distribution 
truncated to a region $A = A_1 \times A_2 \times \cdots \times A_p$ with $A_i$ of the type 
defined in equation \eqref{eq:probz} can be simply obtained by truncating each of the $p$ 
components independently. Draws from a univariate truncated normal can for example be quite 
efficiently obtained via the mixed rejection algorithm of \citet{art:Geweke91} or for truncation near 
the mean by the recent method proposed by \citet{art:Chopin11}.  When a better and much more 
structured guess is available for the initial covariance matrix $\bs{\Sigma}^0$, such 
that the components cannot be truncated independently and the efficient univariate 
methods cannot be applied, the SMC sampling method proposed above is then of course 
a valid alternative.

On the other hand once the initial sample has been obtained (from whichever method of 
choice) sequential Monte Carlo methods provide a natural machinery for efficient parameter updating 
during the E-step of a Monte Carlo based EM. In fact given a particle approximation from the truncated 
target distribution corresponding to the initial parameter values a sequential Monte Carlo approach 
can easily be defined to move between subsequent estimates $\bs{\psi}^m=(\bs{\Sigma}^m,\bs{\beta}^m)$ 
without the need to perform the complete truncation again. The M step of iteration $m$ provides
the newly optimised parameters $(\bs{\Sigma}^{m+1},\bs{\beta}^{m+1})$.  While from the previous E step, 
for each observation $j$, a particle approximation is available from a multivariate normal with 
mean $\bs{X}^j \bs{\beta^m}$ and covariance $\bs{\Sigma}^m$ truncated to the region 
$A^j = A_1^j \times A_2^j \times \cdots \times A_p^j$. One wishes to simply move these particles to 
approximate the truncation to the same region $A^j$ of a multivariate normal with updated 
mean $\bs{X}^j \bs{\beta}^{m+1}$ and covariance $\bs{\Sigma}^{m+1}$.  

Translating the particles could lead to the situation where the mean shift would effectively 
imply moving the sample to a bigger region, which would prevent us from using a simplified version of 
the backward kernel $L_n$ \citep[see section 3.3.2.3 of][]{art:DelMoralDJ2006}.  Instead 
the coordinates of the previous sample can just be rescaled by multiplying them by the diagonal matrix $\bs{D}^{-1}$.  
As long as the scaling factors (elements of $\bs{D}^{-1}$) are all positive, this transformation does not affect 
the truncation region and the mean vector is likewise scaled to $\bs{D}^{-1}\bs{X}^j \bs{\beta}^{m}$.  
By simply choosing $\bs{D}$ to set this equal to the new required mean vector $\bs{X}^j \bs{\beta}^{m+1}$ we have 
a particle approximation with the correct mean and truncation region, but with a covariance matrix 
of $\bs{D}^{-1}\bs{\Sigma}^{m}\bs{D}^{-1}$.  A SMC algorithm can then be applied to target a distribution with 
the new covariance matrix $\bs{\Sigma}^{m+1}$.

Multiple sub-steps might be needed to update $\bs{D}^{-1}\bs{\Sigma}^{m}\bs{D}^{-1}$ to $\bs{\Sigma}^{m+1}$, 
depending on how different the two corresponding targets are.  As the EM algorithm progresses however, 
these two must tend to approach each other and a single step will start to suffice.  For each 
observation $j$ the local (to the EM iteration) initial and final distributions of an artificial sequence 
$\{ \pi_n \}_0^T$ can be defined via $\pi_0 = \textsc{TMN}(A^j, \bs{X}^j \bs{\beta}^{m+1}, \bs{D}^{-1}\bs{\Sigma}^{m}\bs{D}^{-1})$ 
and $\pi_T=\textsc{TMN}(A^j, \bs{X}^j \bs{\beta}^{m+1}, \bs{\Sigma}^{m+1})$ respectively. 
The parameter sequence $\{ \theta_n \}_0^T$ then defines the intermediate targets moves  
such that $\theta_0=\bs{D}^{-1}\bs{\Sigma}^{m}\bs{D}^{-1}$ and $\theta_T =\bs{\Sigma}^{m+1}$, 
and possibly only requires a single step. 

The complete SMC EM procedure is outlined in Algorithm~\ref{alg:SMCEM} of \ref{app:algorithms}.
\subsection{M-step in the multivariate probit: alternative approach to the conditional maximisation step}
\label{sec:mstepmp}
The first step of the conditional maximisation in Section~\ref{sec:2stepcm} can be interpreted as follows.
The expression for the covariance matrix $\hat{\bs{\Sigma}}(\bs{\beta})$ of equation \eqref{eq:sigmaopt} 
maximises the $Q$ function $Q(\bs{\psi},\bs{\psi}^m)$ over $\bs{\Sigma}$ for any value of $\bs{\beta}$. 
Therefore it can be substituted in the expression of $Q(\bs{\psi},\bs{\psi}^m)$ in \eqref{eq:qem} providing
a function which only depends on $\bs{\beta}$
\begin{equation} 
2\hat{Q}(\bs{\beta}, \bs{\psi}^m)  =  -N \log \vert \hat{\bs{\Sigma}} \left(\bs{\beta}\right) \vert -Np ,
\label{eq:qhatem}
\end{equation}
where the simplification of the second term to a constant derives from the fact that the argument 
of the trace reduces to the identity matrix. Finding the value $\tilde{\bs{\beta}}$ which 
maximises \eqref{eq:qhatem} over $\bs{\beta}$ and setting 
$\tilde{\bs{\Sigma}}=\hat{\bs{\Sigma}}(\tilde{\bs{\beta}})$ in \eqref{eq:sigmaopt} 
provides the new parameter $\tilde{\bs{\psi}} = (\tilde{\bs{\beta}}, \tilde{\bs{\Sigma}})$ which maximises the
likelihood. Setting to zero the derivative of \eqref{eq:qhatem} with respect to $\bs{\beta}$ leads to the condition
$\Tr \{\hat{\bs{\Sigma}}^{-1} \rmd  \hat{\bs{\Sigma}} \} =0$. Since $\hat{\bs{\Sigma}} = \bs S$, from
the Monte Carlo expression of $\bs S$ obtained when combining \eqref{eq:qem} and \eqref{eq:expest} it follows that
\begin{equation} \nonumber
\rmd  \hat{\bs{\Sigma}} = d \bs S \simeq -\frac{1}{N} \sum_{j=1}^N { \sum_{k=1}^M W^{j (k)}  
\left[(\bs{X}^j \rmd \bs{\beta} )(\bs{Z}^{j (k)} - \bs{X}^j \bs{\beta} )^{\trans} 
+ (\bs{Z}^{j (k)} - \bs{X}^j \bs{\beta} )(\bs{X}^j \rmd \bs{\beta}
)^{\trans}\right] } .
\end{equation}
and again by the cyclicity of the trace the condition $\Tr \{\hat{\bs{\Sigma}}^{-1} \rmd  \hat{\bs{\Sigma}} \} =0$ 
reduces to the condition in \eqref{eq:diffbeta} with $\hat{\bs{\Sigma}}$ instead of $\bs{\Sigma}$
\begin{equation}
2 \left(\rmd \bs{\beta} \right)^{\trans}\sum_{j=1}^N \sum_{k=1}^M W^{j (k)}  
\left(\bs{X}^j \right)^{\trans} \hat{\bs{\Sigma}}^{-1} (\bs{Z}^{j (k)} - \bs{X}^j \bs{\beta} )  = 0.
\end{equation}
Though $\rmd \hat{\bs{\Sigma}}$ is linear in the components of $\bs{\beta}$, the
inverse matrix $\hat{\bs{\Sigma}}^{-1}$ leads to a system
of coupled higher order polynomial equations.  Solving these is impracticable,
but an iterative scheme over a sequence of points $\tilde{\bs{\beta}}^{n}$ can be followed.

Performing Newton-type iterations would be an option, but when the starting point is 
not too far from the maximum a simpler approximate maximisation can be employed.  
Starting from a point $\tilde{\bs{\beta}}^{n}$ first set $\tilde{\bs{\Sigma}}^{n+1} 
= \hat{\bs{\Sigma}}(\tilde{\bs{\beta}}^{n})$, then separate 
$\hat{\bs{\Sigma}} (\bs{\beta}) = \tilde{\bs{\Sigma}}^{n+1} + \Delta
\tilde{\bs{\Sigma}} (\bs{\beta})$. Recall that $ \log \vert \bs G \vert = \Tr \left\{ \log
\bs G \right\}$ for any matrix $\bs G$ and make the 
approximation $\log(\bs I + \bs G) \approx \bs G$ for $\bs G$ near $\bs 0$
\citep{MatrixCookBook, FunctionsMatrices} to rewrite the $\hat{Q}$ function as
\begin{equation} \nonumber
2\hat{Q}(\bs{\beta}, \bs{\psi}^m) \approx -N \Tr \left\{ \log
\tilde{\bs{\Sigma}}^{n+1} \right\} 
-N \Tr \left\{(\tilde{\bs{\Sigma}}^{n+1})^{-1} \Delta \tilde{\bs{\Sigma}} \right\} -Np.
\label{eq:Qhat}
\end{equation}
where the only term depending on $\bs{\beta}$ is $\Delta \tilde{\bs{\Sigma}}$. 
Since $\rmd \Delta \tilde{\bs{\Sigma}} = \rmd \hat{\bs{\Sigma}}$,
when differentiating with respect to $\bs{\beta}$, maximising the previous expression 
is achieved by solving
\begin{equation} 
0= -N\Tr \left\{ (\tilde{\bs{\Sigma}}^{n+1})^{-1} \rmd 
\hat{\bs{\Sigma}} \right\} \simeq 2\left(\rmd \bs{\beta}\right)^{\trans} \sum_{j=1}^N { \sum_{k=1}^M { W^{j
(k)}  ( \bs{X}^j )^{\trans} (\tilde{\bs{\Sigma}}^{n+1})^{-1} (\bs{Z}^{j (k)} -
\bs{X}^j \bs{\beta}) } } ,
\end{equation}
which is again just \eqref{eq:diffbeta} evaluated at a given $\tilde{\bs{\Sigma}}^{n+1}$.
The solution is of the form in \eqref{eq:betaopt} when replacing $\bs{\Sigma}$ with $\tilde{\bs{\Sigma}}^{n+1}$,
so that the next value of $\tilde{\bs{\beta}}$ can be set
as $\tilde{\bs{\beta}}^{n+1} = \hat{\bs{\beta}}(\tilde{\bs{\Sigma}}^{n+1})$.
For a given starting point of the full parameter vector $\tilde{\bs{\psi}}^{n}$ the covariance matrix 
$\tilde{\bs{\Sigma}}^{n+1}$ and subsequently the coefficient vector $\tilde{\bs{\beta}}^{n+1}$ 
can be updated in this way to provide a point $\tilde{\bs{\psi}}^{n+1}$ for the next iteration.
\subsubsection{Complete maximisation}\label{Mstep}
Because of the logarithmic approximation, each value of $\tilde{\bs{\beta}}^{n+1}$ 
found in a single step does not yet maximise $\hat{Q}$. It is clear however that 
the value of $\bs{\beta}$ can be iteratively adjusted until the 
maximum is found.  The procedure moves along the sequence $\tilde{\bs{\beta}}^{n}$ using 
the current value as the starting point for the next iteration, while
updating $\tilde{\bs{\Sigma}}^{n+1}$ at the same time.  The maximisation can be completed 
by iterating until one finds the maximiser $\tilde{\bs{\psi}}= \lim_{n\to\infty} \tilde{\bs{\psi}}^{n}$, 
and numerically stopping the iterations when the Euclidean norm 
$\vert\vert \tilde{\bs{\beta}}^{n+1} - \tilde{\bs{\beta}}^{n} \vert\vert$ is small.
For the EM algorithm one can then set $\bs{\psi}^{m+1}= \tilde{\bs{\psi}}$.

In general the surety of convergence or even of not decreasing $\hat{Q}$ 
is lost with approximations.  But choosing $\tilde{\bs{\beta}}^{0}=\bs{\beta}^m$ 
(or $\tilde{\bs{\psi}}^{0}=\bs{\psi}^{m}$) as a starting point leads to
the same expression as that found after the first 
conditional maximisation over $\bs \Sigma$ in \eqref{eq:sigmaopt} and $\tilde{\bs \Sigma}^{1} = 
\bs \Sigma^{m+1}$.  The logarithmic approximation then gives $\tilde{\bs{\beta}}^{1}=\bs{\beta}^{m+1}$, 
so that neatly, a single iteration using the logarithmic approximation and the two-step 
conditional maximisation of \cite{art:MengR93} are equivalent when started at 
the same point ($\bs{\psi}^{m}$ for example).  That each iteration (without a particle approximation) 
does not decrease the likelihood follows from the arguments in \cite{art:MengR93}, confirming the 
convergence of the maximisation.  More importantly, completing the maximisation by iterating until 
convergence can equivalently be achieved by running through the two-step conditional maximisation many times.  
\subsection{From generalised EM to EM}
Though the focus of Section~\ref{Mstep} is on multivariate normals, cycling through the conditional 
maximisations of \cite{art:MengR93} until convergence can be applied more generally, 
turning the \emph{generalised} EM of their single round procedure into an EM again.  However, 
as they mention, it may be computationally advantageous to perform an E step between 
conditional maximisations when these are more demanding, and in such cases
the algorithm remains a generalised one.
\subsection{Invariance and identifiability in the multivariate probit model}\label{sec:identifymore}
The full parameter space $\Psi$ of a multivariate probit model comprises $p(p+1)/2$ entries
from the covariance matrix $\bs{\Sigma}$ and $k$ regression coefficients from $\bs{\beta}$. 
Invariance of the likelihood is observed under a rescaling of the coordinates of the latent
multivariate normal variable $\bs Z$ by means of a diagonal matrix $\bs{D}$ with positive 
entries $(d_1, \ldots, d_p)$ according to the transformation $\bs{z}^j = \bs{D}\bs{u}^j$.
The covariance matrix $\bs{\Sigma}$ gets transformed to 
$\bs{\Omega} = \bs{D}^{-1} \bs{\Sigma} \bs{D}^{-1}$
and the vector of regression coefficients $\bs{\beta}$ to 
$\bs{\lambda}= ( d_1^{-1}\bs{\beta}_1^{\trans},\ldots,d_p^{-1}\bs{\beta}_p^{\trans} )^{\trans}$,
but it can easily be checked that the likelihood is left unchanged. 
Choosing the entries of 
$\bs{D}$ to be the square root of the diagonal elements of $\bs{\Sigma}$ 
reduces $\bs{\Omega}$ to correlation form. The invariant space $\Delta$ 
then has coordinates given by the $p$ diagonal elements of $\bs{\Sigma}$ 
(i.e.\ $\delta_1=1/\surd{\sigma_{11}}$ etc.) while the reduced space $\Xi$ includes 
the $p(p-1)/2$ rescaled upper triangular elements of $\bs{\Omega}$ 
(i.e.\ $\omega_{ij}=\delta_{i}\delta_{j}\sigma_{ij}$) and the $k$ elements 
of $\bs{\lambda}=(\delta_1\bs{\beta}_1^{\trans},\ldots,\delta_p\bs{\beta}_p^{\trans})^{\trans}$.  

The likelihood however is not maximised directly, but through the function
\begin{equation} \label{eq:qoriginal}
Q(\bs{\psi}, \bs{\psi}^m) = \sum_{j=1}^N \int_{A^{j}} \textsc{TMN}(A^j,  \bs{X}^j \bs{\beta}^m, \bs{\Sigma}^m) 
\left [ \log \left(\frac{1}{\vert \bs{\Sigma}\vert^{1/2}}\right) 
-\frac{1}{2} (\bs{z}^{(j)}-\bs{X}^j \bs{\beta})^{\trans} 
\bs{\Sigma}^{-1} (\bs{z}^j-\bs{X}^j \bs{\beta})  \right ] \rmd \bs{z}^j .
\end{equation}
Given a diagonal matrix $\bs{D}$ the expression in \eqref{eq:qoriginal} above is only invariant
under a change of the integration variables $\bs{z}^j = \bs{D} \bs{u}^j$, 
if a factor $\vert \bs{D}\vert$ is included inside the log and correspondingly 
inside the log of $H(\bs{\psi}, \bs{\psi}^m)$. 
Moreover, both $\bs{\psi}$ and $\bs{\psi}^m$ need to be scaled by the
same matrix so that essentially $\bs{\delta}=\bs{\delta}^m$. Therefore 
$\bs{\psi}$ and $\bs{\psi}^m$ are tied together in the $Q$ function 
in an apparent constraint, though theoretically they have independent invariant 
spaces for the likelihood.
In \citet{art:ChibG98} maximisation is performed inside the constrained space $\Xi$, while keeping $\delta_i=1$.
Denote by $\bs{\psi}_{\mathrm{c}}$ the corresponding solution
and by $\bs{\psi}_{\mathrm{u}}$ the one obtained through unconstrained maximisation of $Q$.
Clearly $Q(\bs{\psi}_{\mathrm{u}}, \bs{\psi}^m)\geq Q(\bs{\psi}_{\mathrm{c}}, \bs{\psi}^m)$, 
but when projecting $\bs{\psi}_{\mathrm{u}}$ to a point $\bs{\psi}_{\mathrm{p}}$ in the constrained 
space $\Xi$ by setting $\delta_i=1$ then $Q(\bs{\psi}_{\mathrm{p}}, \bs{\psi}^m)\leq Q(\bs{\psi}_{\mathrm{c}}, \bs{\psi}^m)$. 
Since the likelihood is invariant under this projection
\begin{equation} \nonumber
Q(\bs{\psi}_{\mathrm{u}}, \bs{\psi}^m) - Q(\bs{\psi}_{\mathrm{p}}, \bs{\psi}^m) 
= H(\bs{\psi}_{\mathrm{u}}, \bs{\psi}^m) - H(\bs{\psi}_{\mathrm{p}}, \bs{\psi}^m) ,
\end{equation}
and without any information on $H(\bs{\psi}_{\mathrm{u}}, \bs{\psi}^m) - H(\bs{\psi}_{\mathrm{c}}, \bs{\psi}^m)$,  
for example from the second differential of the likelihood as in parameter expanded EM \citep{art:LiuRW98},
it is impossible to say which maximisation increases the likelihood most and is to be
preferred in that respect.
\subsubsection{Reintroducing the likelihood invariance in the $Q$ function}\label{reinvariant}
To remove the above ambiguity, $Q$ can be redefined to respect the invariance of the likelihood, 
for example by replacing the parameters $(\bs{\Sigma},\bs{\beta})$ in \eqref{eq:qoriginal} by their projection 
$(\bs{\Omega},\bs{\lambda})$. Such a replacement effectively enforces invariance of the resulting function 
$\tilde{Q}$ with respect to a rescaling of $(\bs{\Sigma},\bs{\beta})$, making constrained and unconstrained 
maximisation identical.  However, this is no longer true when a (cyclical) two-step 
conditional maximisation is performed. 

With the replacement in \eqref{eq:qem}, $\tilde{Q}$ becomes
\begin{equation} \label{eq:tildeqem}
\tilde{Q}(\bs{\psi}, \bs{\psi}^m)  =  -\frac{N}{2} 
\bigg [ \log \frac{\vert \bs{\Sigma} \vert}{\vert \bs{D} \vert^{2}} + \Tr 
\bigg \{ \bs{D}\bs{\Sigma}^{-1}\bs{D} \tilde{\bs{S}} \bigg \} \bigg ], \qquad
\tilde{\bs{S}} = \frac{1}{N} \sum_{j=1}^N { \sum_{k=1}^M { W^{j (k)}  
(\bs{Z}^{j (k)} - \bs{D}^{-1}\bs{X}^j \bs{\beta} )
(\bs{Z}^{j (k)} - \bs{D}^{-1}\bs{X}^j \bs{\beta} )^{\trans} } } ,
\end{equation}
in terms of the particle approximation in \eqref{eq:expest}, 
and where $\bs{D}$ is a diagonal matrix whose elements are the square roots of the diagonal 
elements of $\bs{\Sigma}$ (so that its projection into correlation form is 
$\bs{\Omega}=\bs{D}^{-1}\bs{\Sigma}\bs{D}^{-1}$). Though $\tilde{Q}$ 
may appear to be limited to the constrained space, it depends on the full parameter 
space when one of $\bs{\Sigma}$ or $\bs{\beta}$ are given. Assume that for given 
$\bs{\psi}^m$ and $\bs{\beta}^{m}=\bs{\lambda}^{m}$ we wish to find $\bs{\Sigma}^{m+1}$. Constrained 
maximisation enforces $\delta_i=1$ to find $\bs{\Omega}_{\mathrm{c}}^{m+1}$.  An unconstrained 
maximisation allows $\delta_i$ to vary, leading to $\bs{\Sigma}_{\mathrm{u}}^{m+1}$  
such that $\tilde{Q}((\bs{\Sigma}_{\mathrm{u}}^{m+1},\bs{\beta}^m), \bs{\psi}^m) 
\geq \tilde{Q}((\bs{\Omega}_{\mathrm{c}}^{m+1},\bs{\lambda}^m), \bs{\psi}^m)$. 
Because of the invariance, the projection of $(\bs{\Sigma}_{\mathrm{u}}^{m+1},\bs{\beta}^m)$ does 
not now change $\tilde{Q}$ resulting in a point in the constrained space with a higher value. 
It can now be unambiguously seen that the unconstrained maximisation is preferable.
In fact $\bs{\beta}$ is only defined up to a scale, which need not be preserved 
during each conditional maximisation, nor given the stochastic nature of the estimation step.
\subsubsection{Constrained maximisation}\label{sec:constrainedmax}
Introducing the invariance of the likelihood into the function $Q$ to obtain the $\tilde{Q}$ in 
\eqref{eq:tildeqem} provides a function whose maximisation allows constrained maximisation 
to be performed over the original $Q$ since they are identical when $\bs{D}$ is the identity matrix.  

First we differentiate \eqref{eq:tildeqem} to obtain the maximisation conditions
\begin{equation} \label{eq:diffQtilde}
2\rmd \tilde{Q} = -N \Tr \left[\bs{\Sigma}^{-1} \left(1-\bs{D}\tilde{\bs{S}} \bs{D}\bs{\Sigma}^{-1} \right)
\rmd \bs{\Sigma}
- 2\bs{D}^{-1} \rmd \bs{D} + \bs{\Sigma}^{-1} \rmd \left(\bs{D} \tilde{\bs{S}} \bs{D} \right) \right]  = 0 ,
\end{equation}
with
\begin{equation} \label{eq:diffDStildeD}
\rmd \left(\bs{D} \tilde{\bs{S}} \bs{D} \right) = \frac{1}{N} \sum_{j=1}^N { \sum_{k=1}^M { W^{j (k)}  
\left[(\rmd \bs{D}\bs{Z}^{j (k)} - \bs{X}^j \rmd \bs{\beta} )(\bs{D} \bs{Z}^{j (k)} - \bs{X}^j \bs{\beta})^{\trans}+ 
(\bs{D}\bs{Z}^{j (k)} - \bs{X}^j \bs{\beta} )(\rmd \bs{D} \bs{Z}^{j (k)} - \bs{X}^j \rmd \bs{\beta})^{\trans} \right]  } } ,
\end{equation}
where $\tilde{\bs{S}}$ now depends on both $\bs{D}$ and $\bs{\beta}$.  

Performing conditional maximisation by fixing $\bs{\Sigma}$ (and hence $\bs{D}$), 
the value of $\hat{\bs{\beta}}$ satisfying equations \eqref{eq:diffQtilde} and \eqref{eq:diffDStildeD} is 
\begin{equation} \label{eq:betaoptQtilde}
\hat{\bs{\beta}} \left(\bs{\Sigma}\right)= 
\bigg ( \sum_{j=1}^N { (\bs{X}^j)^{\trans} \bs{\Sigma}^{-1}  \bs{X}^j }  \bigg )^{-1} 
\sum_{j=1}^N { (\bs{X}^j)^{\trans} \bs{\Sigma}^{-1} \bs{D}
\sum \limits _{k=1}^M {  \left ( W^{j (k)} \bs{Z}^{j (k)} \right ) }  } ,
\end{equation}
which is almost the same as in \eqref{eq:betaopt} but with an extra factor $\bs{D}$ before 
the sum over $k$.

Maximisation with fixed $\bs{\beta}$ over $\bs{\Sigma}$ can in 
turn be done in two steps.  The differential $\rmd \bs{\Sigma}$ is split into a 
diagonal ($2\bs{D}\rmd \bs{D}$) and an off-diagonal part.  The condition for the latter to vanish is 
that $\bs{\Sigma}^{-1}(1-\bs{D}\tilde{\bs{S}} \bs{D}\bs{\Sigma}^{-1})$  be a diagonal matrix, or equivalently 
that $(\bs{\Omega}^{-1}-\bs{\Omega}^{-1}\tilde{\bs{S}}\bs{\Omega}^{-1})$ is the diagonal matrix $\bs{A}$. 
As long as the diagonal elements of $\tilde{\bs{S}}$ are not too 
far from 1, a solution can be found by a simple iterative approach starting from an 
arbitrary $\bs{\Omega}_0$ and then solving for the diagonal matrix $\bs{A}$ the linear equations
\begin{equation}
\bs{\Omega}_{k+1} = \tilde{\bs{S}} + \bs{\Omega}_k \bs{A} \bs{\Omega}_k ,
\label{eq:OmegaA}
\end{equation}
so that $\bs{\Omega}_{k+1}$ is in correlation form. Iterations are repeated until numerical 
convergence provides the required $\bs{\Omega}$. Since for fixed $\bs{D}$ the diagonal part 
of $\rmd \bs{\Sigma}$ is identically zero the steps above allow constrained 
maximisation to be performed for both \eqref{eq:tildeqem} and \eqref{eq:qem}.

The above procedure leads to a significant speed up with respect to numerical optimisation routines
over the off-diagonal elements of $\Omega$. Both methods involve 
inverting and multiplying $p\times p$ matrices, but the relative complexity of the numerical optimisation
would be expected  to grow at least as fast as the number of off-diagonal parameters, namely as 
$p(p-1)/2$ in dimension $p$. In a simple test comparing to the
`nlm' function of the stats package in R the target parameter was set to a noisy version of the 
identity matrix, which was used as starting point for both algorithms. The method here
was over 13 times faster in dimension 4, nearly 40 times faster in dimension 6 and about 100 times 
for $p=8$, highlighting the scale of improvement that can be expected, 
and is consistent with a growth like $p^2$ or better.

If $\bs{D}$ can vary, for the diagonal elements of $\rmd \bs{\Sigma}$ to vanish the matrix
\begin{equation}
\bs{A}-\bs{I}+\bs{\Omega}^{-1}\frac{1}{N}\sum_{j=1}^N \sum_{k=1}^M  W^{j (k)}\bs{Z}^j (\bs{Z}^j)^{\trans}
-\bs{\Omega}^{-1}\bs{D}^{-1}\frac{1}{N}\sum_{j=1}^N \sum_{k=1}^M  W^{j (k)} \bs{X}^j \bs{\beta}
(\bs{Z}^j)^{\trans},
\label{eq:Dmax}
\end{equation}
must have zero along the diagonal. The condition translates into a linear equation in the inverse elements
of $\bs{D}$ and so can likewise be solved easily.  The solution depends on $\bs{\Omega}$, 
which in turn depends (through $\tilde{\bs{S}}$) on $\bs{D}$.  The unconstrained maximisation of 
\eqref{eq:tildeqem} over $\bs{\Sigma}$ for a given $\bs{\beta}$ requires then 
cycling through solving \eqref{eq:Dmax} and \eqref{eq:OmegaA}. As such, the difference 
between constrained and unconstrained maximisation is made transparent.
\subsection{Identifiability for specific formulations of the multivariate probit model}\label{sec:modelconstraints}
\begin{table*}
\caption{Special formulations of the multivariate probit model with a $p$-dimensional response variable $\bs y = (y_1, \ldots, y_p)^\trans$. 
The form of the design matrix with the covariates associated to each observation is provided, where however the observation index $j$ 
is dropped to simplify the notation. The scaling matrix $\bs D$ is defined as $\bs{D} = \mbox{diag}(d_1, \ldots, d_p)$ with $d_i = \sqrt{\sigma_{ii}}$
and $\sigma_{ii}$ the $i$-th diagonal element of $\bs{\Sigma}$. Likelihood invariance means that $\mathcal{L} (\bs{\beta}, \bs{\Sigma}) =  \mathcal{L}(\bs{\lambda}, \bs{\Omega})$ holds under the given transformation of the parameters.
$E(\bs Y) = \Phi_A(\cdot)$ is shorthand for the model definition in \eqref{eq:probz}.}
\begin{tabular}{p{.2\textwidth}p{.25\textwidth}p{.22\textwidth}p{.21\textwidth}}
\hline
& Design matrix & Regression coefficients & Likelihood invariance\\
\hline\noalign{\smallskip}
General form & block diagonal & size $k = \sum_i k_i$ vector & \\[1ex]
$E(\bs Y) = \Phi_A(\bs {X \beta}; \bs{\Sigma})$
&
$\bs{X} = \mbox{diag}((\bs{x}_1)^{\trans}, \ldots, (\bs{x}_p)^{\trans} )$
&
$\bs{\beta} = (\bs{\beta}_1^{\trans}, \ldots, \bs{\beta}_p^{\trans})^{\trans}$
&
$ \bs{\Omega} = \bs{D}^{-1} \bs{\Sigma} \bs{D}^{-1} $
\\
& $ \bs{x}_i = (x_{i1}, \ldots, x_{ik_i})^\trans $
& 
$\bs{\beta}_i = (\beta_{i1}, \ldots \beta_{ik_i})^\trans$
&
$ \bs{\lambda} = \left( d_1^{-1} \bs{\beta}_1^\trans, \ldots, d_p^{-1} \bs{\beta}_p^\trans \right)^\trans $
\\
\hline\noalign{\smallskip}
Shared covariates & size $k$ vector & $p \times k$ matrix  & \\[1ex]
$E(\bs Y) = \Phi_A(\bs {\beta X}; \bs{\Sigma})$ 
&
$\bs{X} = ( x_1, \ldots, x_k )^\trans$
&
$ \bs{\beta} =  (\bs{\beta}_1, \ldots, \bs{\beta}_p)^{\trans}$
& 
$ \bs{\Omega} = \bs{D}^{-1} \bs{\Sigma} \bs{D}^{-1} $
\\
&& $\bs{\beta}_i = (\beta_{i1}, \ldots \beta_{ik})^\trans$
&
$ \bs{\lambda} =  \bs{D}^{-1} \bs{\beta} $
\\
\hline\noalign{\smallskip}
Shared coefficients & $p \times k$ matrix & size $k$ vector & \\[1ex]
$E(\bs Y) = \Phi_A(\bs {X}_{\rm c} \bs{\beta}_{\rm c}; \bs{\Sigma})$
&
$ \bs{X}_{\rm c} =  (\bs{x}_1, \ldots, \bs{x}_p)^{\trans}$
&
$\bs{\beta}_{\rm c} = ( \beta_1, \ldots, \beta_k )^\trans$
&
$ \bs{\Omega}_{\rm c} =  d_1^{-2} \bs{\Sigma} $
\\
&
$\bs{x}_i = (x_{i1}, \ldots, x_{ik})^\trans$
&&
$ \bs{\lambda}_{\rm c} = d_1^{-1} \bs{\beta}_{\rm c} $
\\
\hline\noalign{\smallskip}
\end{tabular}
\label{tab:pbitforms}
\end{table*}
As pointed out in section \ref{sec:identify} the identifiability of the parameters of a model 
is directly related to any invariance of the likelihood. In order to correctly evaluate the identifiability 
of a given model it is then crucial to account for any constraints explicitly or implicitly imposed
on the parameter space. In an attempt to clarify sources of confusion, different formulations
of the multivariate probit models are considered in detail. 
For clarity the different cases are summarised in Table~\ref{tab:pbitforms}.

The most general form of multivariate probit model described in Section~\ref{sec:mvtprobit} 
allows for a different number of covariates for each component of the response variable
and consequently different vectors of regression coefficients. The design matrix for each observation
is block diagonal, with one block, in the form of a row vector, for each response. Depending on the data at hand
the model can be specialised in different ways. The covariates may be shared between the
components of the response variables, while keeping the vectors of regression coefficients different.
This is the case for example in \citet{art:TalhoukDM2012} and \citet{art:XuC2010}. 
Indeed this is simply a special case of the most general formulation since sharing the covariates 
does not change the number of free parameters of the problem, meaning that
the dimension of the invariant space remains the same. Having a single vector of covariates however
allows the problem to be represented in a slight different form, where the regression coefficients can be
packed in a matrix with each row corresponding to a given response component, and the design matrix
reduced to a single column vector. The parameter transformation yielding invariance of the likelihood 
can then be written in a more compact form, as summarised in Table~\ref{tab:pbitforms}. 
An important practical consequence is that a closed form solution can be derived for the M-step 
as pursued in \citet{art:XuC2010}, provided that no constraints are imposed.

Alternatively the regression coefficients, and the number of covariates, may be shared among the responses.
The model chosen in \cite{art:ChibG98} for the Six Cities dataset is of this type, the same number of covariates
are observed for each component of the response variables, but they take different values. 
The model can be represented into a more compact form with a $p \times k$ design matrix $\bs{X}_{\rm c}$, where 
each row corresponds to one component of the response variable and a $k$-dimensional vector of 
regression coefficients $\bs{\beta}_{\rm c}$ (see Table~\ref{tab:pbitforms}).
The conditional maximisation in Section~\ref{Mstep} can be applied to maximise 
over the constrained space of $(\bs{\Sigma},\bs{\beta}_{\rm c})$. The extreme case of fixing both the 
covariates and the regression coefficients would lead to a univariate probit model. 

Fixing the vector of regression coefficients across response components accounts to imposing 
constraints at the modelling stage, with the effect of reducing the number of free parameters and 
hence the dimension of the invariant space. This aspect seems to have been overlooked in 
the literature where the Six Cities dataset is taken as an example, including \cite{art:ChibG98}. 
It is simply treated as a special case of the most general form, and a correlation structure 
is imposed on the covariance matrix, but in fact this is unnecessary for identifiability.

Consider the standard formulation where the design matrix $\bs{X}$ is block diagonal, with each element
of the same length $k$, and the vector of regression coefficients is a vector of length $p \times k$ with $p$ 
repeated sub-vectors $\bs{\beta}_{\rm c}$. It has been noted in Section~\ref{sec:identifymore} that the transformation 
which guarantees invariance is such that each subvector $\bs{\beta}_i$ is multiplied by a 
different positive factor $d_i$, violating the desired constraint that they be all equal.
In order to avoid that the sub vectors of regression coefficients differ
they all need to be multiplied by the same factor. In other words the likelihood is now 
left unchanged, independently of $\bs{X}^{j}$, only when rescaling all the coordinate directions 
by the same amount, corresponding to a one dimensional invariant space. A reduced space 
can be defined by fixing the first diagonal element of the covariance matrix to 1, 
call $(\bs{\Omega}_{\rm c},\bs{\lambda}_{\rm c})$ the corresponding parameters. An invariant $\tilde{Q}$ 
is obtained by replacing $\bs{X}^{j}$,$\bs{\Sigma}$ and $\bs{\beta}$ in \eqref{eq:qoriginal} 
by $\bs{X}_{\mathrm{\rm c}}^{j}$, $\bs{\Omega}_{\rm c}$ and $\bs{\lambda}_{\rm c}$ respectively and 
by setting all the elements of $\bs{D}$ in \eqref{eq:tildeqem} to be the square root of the 
first element of $\bs{\Sigma}$.  Constrained and unconstrained maximisation follow from 
the considerations in Section~\ref{sec:identifymore} but with the slight changes that only 
the first element of the matrix $\bs{A}$ in \eqref{eq:OmegaA} 
is non-zero and just the trace of \eqref{eq:Dmax} needs to be 0.

Maximising over an overly constrained space leads in general to a 
lower likelihood than when only imposing the conditions needed to ensure identifiability.  
Nevertheless, were the correlation form desired for modelling reasons, maximisation 
can be performed by setting $\bs{D}$ to be the identity matrix and using $\bs{X}_{\mathrm{c}}^{j}$ 
in the formulae \eqref{eq:betaoptQtilde} and \eqref{eq:OmegaA} above.

\section{Results}
\subsection{Comparison of the SMC and Gibbs samplers for truncated multivariate normals}
The SMC method for sampling truncated multivariate normals is now compared to a Gibbs sampler \citep{art:Geweke91, art:Robert95} which is a Markov Chain where each component is sampled conditional on all the others. In dimension $p$ each `pass' of the Gibbs sampler requires drawing $p$ univariate truncated normal variables which can be efficiently achieved by the mixed accept-reject algorithm of \cite{art:Geweke91} or for better efficiency near the mean using the tabulated accept-reject of \cite{art:Chopin11}. The Gibbs sampler starts from the correct truncation region but, as noted in \cite{art:Geweke91}, it can be rather slow at converging to the correct correlation structure. Convergence however is improved for extreme truncations since these have the effect of reducing the correlation among the components.

The SMC sampler on the other hand starts with the `correct' correlation structure and moves to the required truncation region. Better performance could then be expected for correlated samples with the Gibbs sampler becoming preferable for more extreme truncation or lower correlation. A simulation study is conducted for the type of truncation regions that occur for the multivariate probit model in four dimensions, $p=4$. The binary variables are set to $y_i=1$, corresponding to the quadrant with positive $\bs{z}$, and the vector of means is chosen as $\bs{\mu}=(-1,-1,1,1)^\trans$ so that the mean is included in two dimensions and excluded in the other two.  All the off-diagonal elements of the correlation matrix $\bs{\Sigma}$ are set equal to $\rho$.  The matrix of second moments is estimated using 10 million samples obtained by rejection sampling. Estimates are then obtained from the SMC and the Gibbs samplers, and a statistic $F$ is defined as the square root of the mean square distance between each estimate and the reference value from the rejection sampling.  This process is repeated for a range of $\rho$ values.
\begin{figure}
\begin{center}
  \includegraphics[width=0.6\textwidth]{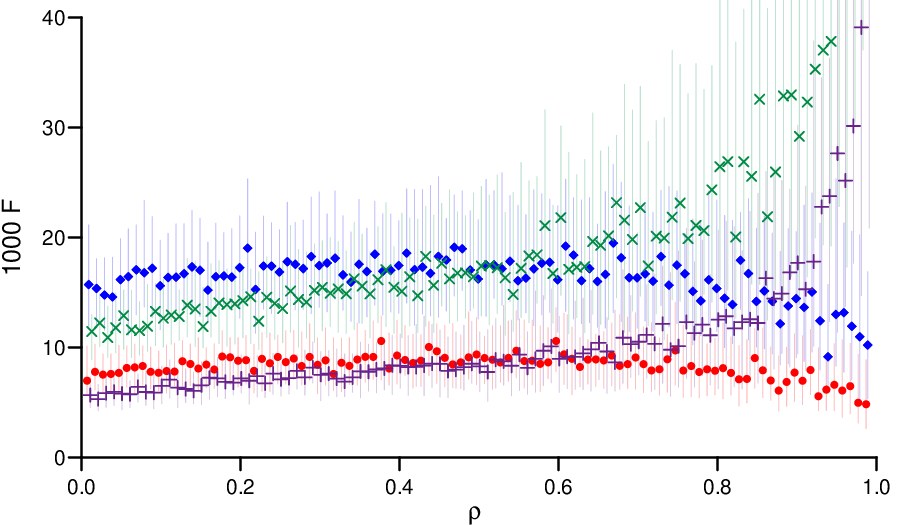}
\end{center}
\caption{The median error $F$ in estimating the matrix of second moments in 4d as well as the inter-quartile ranges for the SMC sampler (diamonds) and a Gibbs sampler (crosses).  Also included are runs with 4 times as many particles for the SMC (dots) and Gibbs (plusses) sampler.}
\label{fig:4dcomparison}
\end{figure}

Initially both algorithms are run to obtain 10\,000 samples; for the Gibbs sampler this is the sample size after discarding the first fifth as burn in.  For each value of $\rho$ we run both samplers 100 times and plot the median value of $F$ as well as the inter-quartile range in Fig.~\ref{fig:4dcomparison}.  The error in the Gibbs sampler increases with $\rho$ and starts to increase rapidly after $\rho\approx0.75$.  The error from the SMC sample on the other hand is fairly constant and actually starts to decrease for large $\rho$.  Both samplers then have similar performance for moderate correlations with the SMC approach notably better for large correlations.  In the central range of $\rho\approx 0.5$ the truncation actually reduces the correlation so that the off-diagonal elements of the sample correlation matrix are only around $0.23$ on average.

However the Gibbs sampler implementation is faster than the SMC version so that a Gibbs chain of about 50\,000 passes can be obtained in the same time as the SMC sampler provides a sample of 10\,000.  Discarding the first fifth leaves a sample which is four times larger and whose error is roughly halved correspondingly.  The growth in the error with $\rho$ still allows the SMC sampler to perform better, but now only for rather large values of $\rho$ above about $0.85$, or where the actual sample correlations are above the more moderate value of approximately $0.56$. Of course the actual computational time depends upon how efficiently each algorithm is implemented, but the simulation here suggests that the SMC sampler will have an advantage for high correlations.

Finally, SMC samples of size 40\,000 are also obtained.  The errors can again be observed to halve compared to the samples of 10\,000.  For both samplers in Fig.~\ref{fig:4dcomparison} the errors observed with the larger sample sizes resemble a scaled version of those from smaller sample sizes.

\subsection{Application: the Six Cities dataset}
To test the validity of our method, we treat the widely analysed data set 
from the Six Cities longitudinal study on the health effects of air pollution, for 
which a multivariate probit model was considered for a range of covariance structures by \cite{art:ChibG98}, who 
conducted both Bayesian and non-Bayesian analysis. Later \cite{art:SongL2005} 
proposed a confirmatory factor analysis for the same model, while more recently \cite{art:Craig2008} 
used the example as a test case for a new method for geometrically 
reconstructing multivariate orthant probabilities which leads to an efficient evaluation of 
probabilities of the type in \eqref{eq:probz}. As opposed to the MCMC procedure 
of \cite{art:ChibG98}, the SMC method provides estimates of the orthant 
probabilities as a by-product of the sampling, so that likelihoods are readily available
for comparison to the results in \cite{art:Craig2008}. Moreover the SMC sampler produces
a sample from the fitted distribution which is useful for further evaluation of 
intractable expectations of interest.

The Six Cities study was meant to model a probabilistic relation over time
between the wheezing status of children, the smoking habit of their mother
during the first year of observation and their age. In particular the subset of
data considered for analysis refers to the observation of $537$ children from
Steubenville, Ohio. The wheezing condition $y_i^j$ of each child $j$ at age
$i \in \{7,8,9,10 \}$ and the smoking habit $h^j$ of their mother are recorded 
as binary variables, with value 1 indicating the condition (wheezing/smoking) present.
Three covariates are assumed for each component $i$, namely the age $x_{i1}^j = i-9$ 
of child $j$ centred at $9$, the smoking habit $x_{i2}^j = h^j$ and an interaction 
term $x_{i3}^j = (i-9) h^j$ between the two.  A probit model can then be constructed 
\begin{equation} \nonumber
\mathrm{pr} \{ y_i^j = 1 \} = \mathrm{pr} (z_i^j > 0) = \Phi\left[(\beta_0 + \beta_1
\cdot x_{i1}^j + \beta_2 \cdot x_{i2}^j + \beta_3 \cdot x_{i3}^j)\sigma_{ii}^{-\frac{1}{2}}\right] ,
\end{equation}
where $z_i^j$ is the $i$th component of a multivariate random variable
$\bs{Z}^j \thicksim \mathcal{N}(\bs{X}_{\mathrm{c}}^j \bs{\beta}, \bs{\Sigma})$ and $\Phi$ is the
cumulative distribution function of a standard normal random variable.

Note that this is an example of a compact model as discussed in 
Section~\ref{sec:modelconstraints} and the invariant space is therefore 
only 1-dimensional. For identifiability it is then sufficient to fix only one of the diagonal elements
of the covariance matrix, which in previous approaches has been 
overly restricted to be in correlation form instead. 
For comparison we therefore first perform 
constrained maximisation with $\bs{\Sigma}$ in correlation form using the methods in 
section~\ref{sec:constrainedmax} and then run the SMC EM algorithm with the correct 
invariance.
\subsubsection{Variance reduction}
To reduce the variance associated with the stochastic nature of the Monte Carlo E step, 
the parameter can be updated according to a stochastic approximation type rule
\begin{equation} \nonumber
\bs{\psi}^m = \bs{\psi}^{m-1} + \zeta_m (\hat{\bs{\psi}}^m - \bs{\psi}^{m-1}) \equiv
(1-\zeta_m)\bs{\psi}^{m-1} + \zeta_m \hat{\bs{\psi}}^m ,
\end{equation}
where $\hat{\bs{\psi}}_m$ is the actual estimate obtained from the M-step and 
$\zeta_m \in (0,1)$ a stepsize with the purpose of gradually shifting the 
relative importance from the innovation $(\hat{\bs{\psi}}^m - \bs{\psi}^{m-1})$ to the
value of the parameter $\bs{\psi}_{m-1}$ learnt through the previous iterations, and 
which therefore goes to 0. 
The scheme is like taking a weighted average of the previous estimates, so 
we refer to it as a `variation reduction' step. This way the monotonicity 
property of the EM algorithm is not guaranteed, but as long as the parameters
remain within a neighbourhood of the maximum likelihood point where it can be 
approximated quadratically, monotonicity trivially follows from the convexity, 
so that in many practical cases this matter may not cause any issues.
\subsubsection{Comparison to alternative approaches with covariance matrix in correlation form}\label{sec:sixCtests}
\begin{table*}
\caption{Maximum likelihood estimates for the Six Cities dataset as obtained by
using the constrained SMC algorithm with linearly increasing number of particles, after 
variance reduction and for a single run where the samples are recycled.  Included for
comparison are the results of \cite{art:ChibG98} and \cite{art:Craig2008}.
The value in brackets next to each estimate is the estimated standard error. 
The values of the parameters (and their errors) have all been multiplied by
1000. The last lines report the estimated $l(\bs{\psi})$ and corrected $\hat{l}(\bs{\psi})$ 
log-likelihoods, and a more accurate value obtained by numerical integrations.}
\label{tab:sixcitiesMLE}
\begin{tabular}{lrlrlrlrlrl}
\hline\noalign{\smallskip}
\multicolumn{3}{c}{\cite{art:ChibG98}}  &
\multicolumn{2}{c}{\cite{art:Craig2008}} & 
\multicolumn{2}{c}{linear increase} & 
\multicolumn{2}{c}{variance reduction} &
\multicolumn{2}{c}{recycled samples} \\
\noalign{\smallskip}\hline\noalign{\smallskip}
$\beta_0$ & -1118 & (65) & -1122 & (62) & -1122 & (62) & -1123 & (62) & -1122 & (62) \\
$\beta_1$ & -79 & (33) & -78 & (31) & 79 & (31) & -79 & (31) & -78 & (31) \\
$\beta_2$ & 152 & (102) & 159 & (101) & 159 & (101) & 159 & (101) & 158 & (101) \\
$\beta_3$ & 39 & (52)  & 37 & (51) & 37 & (51) & 38 & (51) & 37 & (51) \\
$\sigma_{12}$ & 584 & (68) & 585 & (66) & 583& (66) & 583 & (66) & 581 & (66) \\
$\sigma_{13}$ & 521 & (76) & 524 & (72) & 522 & (71) & 522 & (71) & 523 & (71) \\
$\sigma_{14}$ & 586 & (95) & 579 & (74) & 577 & (74) & 578 & (74) & 579 & (73) \\
$\sigma_{23}$ & 688 & (51) & 687 & (56) & 686 & (56) & 686 & (56) & 682 & (57) \\
$\sigma_{24}$ & 562 & (77) & 559 & (74) & 558 & (74) & 558 & (74) & 558 & (74) \\
$\sigma_{34}$ & 631 & (77) & 631 & (67) & 626 & (67) & 627 & (67) & 625 & (67) \\
$l(\bs{\psi})$ & -794$\cdot$94 & (0$\cdot$69) & -794$\cdot$93 & (0$\cdot$66) 
& -794$\cdot$91 &  (0$\cdot$59)& -794$\cdot$95 & (0$\cdot$82) &
-794$\cdot$86 & (0$\cdot$66)\\
$\hat{l}(\bs{\psi})$ & -794$\cdot$70 & & -794$\cdot$72 & & -794$\cdot$73
& & -794$\cdot$61 & &-794$\cdot$65 & \\
& \multicolumn{2}{c}{\{-794$\cdot$749\}}  & 
\multicolumn{2}{c}{\{-794$\cdot$738\}}  & 
\multicolumn{2}{c}{\{-794$\cdot$742\}}  & 
\multicolumn{2}{c}{\{-794$\cdot$740\}}  & 
\multicolumn{2}{c}{\{-794$\cdot$748\}}
 $(10^{-5})$\\
\noalign{\smallskip}\hline 
\end{tabular}
\end{table*}

To fit the model, a SMC sampler is implemented with the number of particles 
increasing linearly from 50 to 2000, over 40 iterations, followed by 10 further steps 
of variance reduction with 4000 particles. As for the tuning parameters of the algorithm 
the desired acceptance probability is set to $\alpha^{\star}=.6$ and the fraction $s$ 
defining the resampling threshold \textsc{ESS}$^{\star}$ as $s=.8$.
Results for the constrained maximisation are presented in
Table~\ref{tab:sixcitiesMLE} along with those of \cite{art:ChibG98} and \cite{art:Craig2008}. 
Good agreement can be observed both for the estimates and the standard errors. The latter
are the square roots of the diagonal elements of the inverse observed Fisher information matrix, 
which in the case of missing data can be obtained from Louis' method \citep{art:Louis82}.

Also given in Table~\ref{tab:sixcitiesMLE} are average values of the corresponding 
log-likelihoods together with the standard deviation estimates over 50 runs. No real differences can be seen, 
with likelihoods comparable to, but slightly below, the estimate 
of -794$\cdot$74 in \cite{art:Craig2008}. Due to the sampling noise the log-likelihood 
tends to be underestimated. A simple correction, discussed in \ref{logcorrection}, 
consisting in adding half the variance over the runs, brings the estimates, 
$\hat{l}(\bs{\psi})$ in Table~\ref{tab:sixcitiesMLE}, closer to that 
in \cite{art:Craig2008}.

In the case of the Six Cities dataset the design matrix $\bs{X}_{\mathrm{c}}$ only takes
two possible values, corresponding respectively to a smoking and non-smoking mother. 
Numerical calculation of the likelihood requires 32 regions to be evaluated for each of the 
parameter value $\bs{\psi}$ estimated by the different methods. Numerical integration in dimension
4 is feasible and gives the results in curly brackets in Table~\ref{tab:sixcitiesMLE}, 
with accuracy $10^{-5}$, confirming that the SMC EM finds better parameter values
than \cite{art:ChibG98}. We have also noticed that the estimates
from other MCMC methods, such as in \cite{art:SongL2005}, seem to be closer to 
those in \cite{art:ChibG98}, while ours are closer to the 
results of the exact method of \cite{art:Craig2008}.

Regardless of the method used for drawing the samples, 
our approach avoids the computationally expensive 
constrained maximisation employed by \cite{art:ChibG98} by replacing it with the simple procedure in
section~\ref{sec:constrainedmax}. Despite the advantage with respect to standard numerical optimisation,
the cost of either method is not significant compared to the sampling time. However failing to take 
advantage of the cyclicity of the trace would make the numerical optimisation substantially more involved.
The evaluation of $Q$ would require quadratic forms as in \eqref{eq:quadforms} to be recalculated for
each value of $\bs{\Sigma}$ during the numerical routine, with an $\mathcal{O}(N)$ increase in cost. 
This might be the reason behind the suggestion at the beginning of page 355 of \cite{art:ChibG98}
to redraw the latent variables between the two conditional maximisations for efficiency reasons.

Assuming that cyclicity of the trace is indeed exploited, a more significant gain comes instead from
completing the maximisation by cycling through the conditional maximisations until convergence, 
rather then performing a single cycle. 
Completing the maximisation in fact reduces the number of full EM iterations needed.

A further major benefit of the SMC method is that the particle approximation can be updated
after each M step and need not necessarily be resampled at each iteration, as described in section~\ref{sec:smcem}.
The last column of Table~\ref{tab:sixcitiesMLE} shows results obtained when recycling the samples in a SMC EM algorithm
with 2000 particles and 40 iterations. Since oscillations before the variance reduction step are around 0$\cdot$001 between 
iterations (with 2000 particles), parameter estimates when recycling the sample are 
essentially equivalent, at a much reduced computational cost. Updating the particle approximation
is about 15 times faster than drawing the sample again, and the entire procedure proves to be 
about 5 times faster than a run with a linear increase of the number of particles drawn
from scratch each time, with the latter strategy still enjoying a factor two improvement over 
keeping the number fixed at 2000. 
Similar parameter values are obtained when using fewer particles, but obviously with higher variance.

\subsubsection{Unrestricted model}
Strictly speaking the Six Cities model does not require $\bs{\Sigma}$ 
to be in correlation form for identifiablity reasons. As discussed in Section \ref{sec:modelconstraints}
the invariant space is in fact only one-dimensional. 
To illustrate an application of the ideas presented in Section \ref{sec:identifymore} 
a more general model which does not impose correlation form is analysed.
To respect the invariance, one can either fix the first element of $\bs{\Sigma}$ 
(i.e.\ set $\sigma_{11}=1$) or run unconstrained maximisation (and project the results).
Unconstrained maximisation was performed over 60 iterations with 4000 particles
before the variance reduction step, since it may take longer for the EM algorithm to 
explore a larger space. A fairly robust point is found with the non-invariant $Q$, while 
the invariant $\tilde{Q}$ seems to lead to a flatter likelihood neighbourhood, with the 
solution appearing more sensitive to the number of particles during earlier iterations or 
on imposing the constraint of fixing $\sigma_{11}$ to 1. Results are given in 
Table~\ref{tab:sixcitiesMLEunconstrained}, and again can be quite closely reproduced by 
recycling the samples in a sequential manner between parameter updates.  
Numerical integrations with $10^{-5}$ accuracy produces the values $-792.849, -792.836, -792.834$,
when working respectively with $Q$, the invariant $\tilde{Q}$ or fixing $\sigma_{11}=1$.
For the latter two, despite the different parameter estimates, the likelihoods are essentially identical, 
and interestingly also higher than the one obtained for the non invariant $Q$. This seems practical
evidence for an advantage in targeting the likelihood more directly through utilising the invariance.

Since there are only two possible forms for the design matrix $\bs{X}_{\mathrm{c}}$, 
a local symmetry arises, along with the global one, when $\beta_0\beta_3=\beta_1\beta_2$.
Moving near this symmetry may allow the EM algorithm to find different final maxima and 
explain the different parameter values found by using invariance or not. The local symmetry 
along with the estimation noise may be responsible for the non positive definiteness of the 
observed Fisher information (resulting from the difference of two positive definite matrices).
Fixing a further parameter value, such as another diagonal element of $\bs{\Sigma}$ to be 1,
removes the local symmetry and allows standard errors to be obtained, centred around $.10$ 
and ranging from $.045$ to $.16$.

\begin{table*}
\caption{Example maximum likelihood estimates for the Six Cities dataset obtained 
using the unconstrained SMC algorithm for non-invariant $Q$, invariant
$\tilde{Q}$ and by fixing $\sigma_{11}=1$.  The standard deviations of the 
log-likelihood estimates are 0$\cdot$90, 0$\cdot$75 and 0$\cdot$70 respectively,
so that the corrected values of the likelihood $\hat{l}(\bs{\psi})$ are -792$\cdot$97,
-792$\cdot$87, -792$\cdot$825 and the numerical values with $10^{-5}$ 
accuracy are -792$\cdot$849, -792$\cdot$836, -792$\cdot$834.  
The values of the parameters have all been multiplied by 1000.}
\begin{tabular}{crrrrrrrrrrrrrr}
\hline\noalign{\smallskip}
& $\beta_0$ & $\beta_1$ & $\beta_2$ & $\beta_3$ & $\sigma_{12}$ & $\sigma_{13}$ 
& $\sigma_{14}$ & $\sigma_{22}$ & $\sigma_{23}$ & $\sigma_{24}$ & $\sigma_{33}$ 
& $\sigma_{34}$ & $\sigma_{44}$ & $l(\bs{\psi})$ \\
\noalign{\smallskip}\hline\noalign{\smallskip}
$Q$ &  -1176 & -84 & 159 & 41 & 647 & 592 & 572 & 1208 & 855 & 619 & 1255 & 715 & 1001 & -793$\cdot$37\\
$\tilde{Q}$ & -1235 & -113 & 168 & 47 & 664 & 622 & 612 & 1275 & 921 & 683 & 1383 & 802 & 1146 & -793$\cdot$15 \\
fixed $\sigma_{11}$ & -1241 & -116 & 169 & 48 & 666 & 626 & 615 & 1279 & 927 & 686 & 1395 & 809 & 1158 & -793$\cdot$07\\
\noalign{\smallskip}\hline 
\end{tabular}
\label{tab:sixcitiesMLEunconstrained}
\end{table*}
\subsection{Higher dimensional simulated dataset}
A higher dimensional example with simulated data is presented to show that the method scales reasonably well.
The model chosen has the same formulation as the one used for the Six Cities case, corresponding to the third line in Table~\ref{tab:pbitforms}. 
The response variable is $8$-dimensional with 7 covariates (including the intercept) associated to each component, resulting in a $8\times7$
design matrix. The entries different from the intercept are drawn from a uniform distribution on the interval $(-.5,.5)$. The parameters are set to
\begin{equation*}
 \bs{\beta_c} = 
 \left(
 \begin{array}{r}
 1.00 \\ 
 0.30 \\ 
-0.30 \\ 
 0.20 \\ 
-0.20 \\ 
 0.10 \\ 
-0.10 \\ 
 \end{array}
 \right),
 \bs{\Sigma} = 
\begin{pmatrix}
 1.00 & 0.10 & 0.10 & 0.10 & 0.10 & 0.20 & 0.20 & 0.40 \\ 
 0.10 & 1.20 & 0.10 & 0.10 & 0.10 & 0.20 & 0.30 & 0.40 \\ 
 0.10 & 0.10 & 1.20 & 0.10 & 0.20 & 0.20 & 0.30 & 0.40 \\ 
 0.10 & 0.10 & 0.10 & 1.10 & 0.20 & 0.20 & 0.30 & 0.50 \\ 
 0.10 & 0.10 & 0.20 & 0.20 & 1.10 & 0.20 & 0.30 & 0.50 \\ 
 0.20 & 0.20 & 0.20 & 0.20 & 0.20 & 0.90 & 0.40 & 0.60 \\ 
 0.20 & 0.30 & 0.30 & 0.30 & 0.30 & 0.40 & 0.90 & 0.60 \\ 
 0.40 & 0.40 & 0.40 & 0.50 & 0.50 & 0.60 & 0.60 & 0.80 \\ 
\end{pmatrix}
\end{equation*}
and 1000 observations are generated from the resulting model. 
Inference is then performed using both our SMC EM method and a
Gibbs based MCEM approach. 
Since, as noted in Section~\ref{sec:sixCtests}, the cost of the M step
with respect to the E step is relatively low in both cases, 
we retain our implementation of the M step rather than 
using numerical optimisation routines, despite marginally
penalising the SMC EM algorithm in the comparison. 
The number of particles M is set to 4000 for both the SMC and the
Gibbs sampler, and 40 iterations of the EM are performed.
The square root of the mean squared distance of the estimated parameters from the real ones
are found to be $.079$ and $.080$. However the distance between them is about $.007$, 
but the two clouds from different runs are barely distinguishable, since they have variations
around .01 within them. Local symmetries are excluded in the simulated example
so the standard errors could be estimated, ranging between about $.045$ and $.172$, 
and centred at $.074$ for the SMC, 
with similar values for the parameter values estimated via Gibbs, and in agreement
with the actual distance to the real values.
The SMC sampler has the advantage over Gibbs 
of automatically providing estimates of the likelihoods. Over $50$ runs
the average log-likelihoods were found to be around $-3280.3$ and $-3280.4$, 
with a standard deviation of $.4$. Although the likelihood for the 
parameters estimated via the SMC EM method happens to be marginally better in this case, 
the noise is too high for an accurate comparison and standard numerical integration is not
an option due to the high dimensionality of the problem. A computational advantage still remains, 
since with the same number of particles the run time for the SMC EM is about 
two and a half times shorter than for the Gibbs based EM.
\section{Conclusions}
A new method based on sequential Monte Carlo samplers is introduced for 
the maximum likelihood estimation of multivariate probit models. In particular an 
adaptive sequential Monte Carlo algorithm is proposed to sample from high dimensional 
truncated normals.  The proposal builds upon the property that a Student $t$ distribution
approaches a Gaussian as the degrees of freedom go to infinity. When comparing to a Gibbs 
sampler the quality of the sample produced by the SMC sampler seems to be better for high 
correlation.

The typical iterative procedure of the EM algorithm appears like the ideal setting for SMC methods,
which provide a natural machinery to evolve the particle approximation from one iteration to the next
when updating the parameters in the M step. Performing the truncation to the current region starting 
from scratch can so be avoided. This way the computational cost is greatly reduced, with no 
particular loss in the performance, as seen for the example of the
Six Cities dataset where similar parameter estimates are obtained when restarting 
at each iteration and with the fast sequential updating scheme. 

Since by construction the sequential Monte Carlo sampler also provides samples from 
truncated Student distributions, it is clear that the method can be easily extended to 
a scenario where a Student $t$ distribution is assumed for 
the underlying latent variable of the probit model, rather than a normal distribution. 
Extensions to models with multinomial response variables are of course also possible.

Furthermore some of the confusion that has arisen around the maximisation step is 
clarified and the first complete EM algorithm for multivariate probit models is presented.
Previously, methods typically proposed in the literature have inevitably resulted 
in a generalised EM, while here the full maximisation is both easy to implement and 
efficient, with almost no computational cost.  By examining the identifiability of such 
models we show that there is in fact a simple way to perform constrained maximisation,
a process which is normally more computationally demanding. More importantly, we 
demonstrate how to tweak the EM algorithm so that it more directly targets increasing 
the likelihood. This is achieved by mimicking the invariance of the likelihood in the 
function $Q$ at the basis of the maximisation process, a strategy that should be of 
interest for other models.

An interesting alternative to EM for point estimation in the context of latent variable models, 
when neither the E step nor the M step are analytically tractable, is provided by a set of methods
combining multiple imputation and simulated annealing ideas, as in \cite{art:DoucetGR2002, art:GaetanY2003, art:JohansenDD2008}.
Sampling is then performed not only in the E step to impute the latent variables, but also in the M step
to draw parameter values which are expected to converge to the maxima of the object function of interest. 
A desirable property of algorithms based on a stochastic version of the M step with respect to its deterministic counterpart
is that they have a chance to escape local maxima. 
Obtaining multiple copies of the latent variables is essentially equivalent to drawing a sample in the
E step of a standard Monte Carlo EM algorithm, therefore the same sampler can be applied. 
In the case of multivariate probit models then the SMC sampler of Section~\ref{sec:smctmn} would also be an option for
the multiple imputations. Drawing the parameters to mimic the M step on the other hand may be non trivial, 
especially in higher dimensions. The difficulties lie in particular with ensuring that the identification constraints
on the covariance matrix are met, as already noted by \cite{art:ChibG98}, and further discussed for example
by \cite{art:McCullochPR2000, art:Nobile2000}, in relation to multinomial probit models. More recently a 
parameter expanded method to simulate correlation matrices has been suggested by \cite{art:LiuD2006}.
However in the context of multivariate probit models we show that performing the M step is actually pretty straightforward.
\section*{Acknowledgements}
The authors would like to thank the associate editor and two referees for their comments
and suggestions which greatly helped to improve the overall presentation and readability of the manuscript,
and thank Quirin Hummel for implementing the numerical integrations for the likelihoods of the Six Cities example.
\appendix
\section{The $Q$ function for the probit model}\label{app:QfunProbit}
For the multivariate probit model, substituting \eqref{eq:completelike} into the expectation in \eqref{eq:Qfun} gives
\begin{eqnarray}
\nonumber Q(\psi, \psi^m) &=& \mathbb{E}_{\bs{Z} \vert \bs{Y}, \psi^m} \left[ l(\psi \vert \bs{Y,Z}) \right]
= \int_{\bs{z} \vert \bs{y}, \psi^m} \sum_{j=1}^N {\log \left [ I_{A^j}(\bs{z}^j) \phi(\bs{z}^j; \bs{X}^j \bs{\beta}, \bs{\Sigma}) \right ]} \cdot \prod_{l=1}^N \pi(\bs{z}^l \vert \bs{y}^l, \psi^m) \de \bs{z}^1 \cdots \bs{z}^N \\ 
&&
\label{eq:Qfun1}
\end{eqnarray}
where $ \phi(\bs{z}^j; \bs{X}^j \bs{\beta}, \bs{\Sigma})$ is the density of a multivariate normal distribution and $\pi(\bs{z}^l \vert \bs{y}^l, \psi^m)$ is the density of a truncated multivariate normal constrained to the domain $A^l$. Denote it by $\mbox{TMN}(\bs{z}^l; A^l,  \bs{X}^l \bs{\beta^m}, \bs \Sigma^m)$ in the following.
After inverting the order of integration and summation, and accounting for the fact that the integrals with respect to all the variables $\bs{z}^l$ for $l \neq j$ can be independently evaluated and are normalised
\begin{equation*}
\int_{ \substack{\bs{z}^l \vert \bs{y}^j, \psi^m \\ l \neq j} } \limits \prod_{\substack{l=1 \\ l \neq j} } \mbox{TMN}(\bs{z}^l; A^l,  \bs{X}^l \bs{\beta^m}, \bs{\Sigma}^m)  \prod_{\substack{l=1 \\ l \neq j} } \de \bs{z}^l =1,
\end{equation*}
the integral in \eqref{eq:Qfun1} then simplifies to
\begin{eqnarray} \label{eq:Qfun2}
Q(\psi, \psi^m) = \sum_{j=1}^N \int_{\bs{z}^j \vert \bs{y}^j, \psi^m} \log \left [ I_{A^j}(\bs{z}^j) \phi(\bs{z}^j; \bs{X}^j \bs{\beta}, \bs{\Sigma}) \right ] \mbox{TMN}(\bs{z}^j, A^j,  \bs{X}^j \bs{\beta^m}, \bs{\Sigma}^m) \de \bs{z}^j,
\end{eqnarray}
with $I_{A^j}(\bs{z}^j) \equiv 1$ on the domain of integration since $\mbox{TMN}(A^j,  \bs{X}^j \bs{\beta}^m, \bs \Sigma^m)$ is only different from zero for $\bs{z}^j \in A^j$. Substituting into \eqref{eq:Qfun2} the expression for the density of a multivariate Gaussian density, and neglecting the proportionality constant term which is irrelevant for the maximisation,
\begin{equation} \nonumber
\phi(\bs{z}^j; \bs{X}^j \bs{\beta}, \bs{\Sigma}) \propto \vert \bs{\Sigma} \vert^{(-1/2)} \exp \left(-\frac{1}{2} (\bs{z}^{(j)}-\bs{X}^j \bs{\beta})' \bs{\Sigma}^{-1} (\bs{z}^j-\bs{X}^j \bs{\beta}) \right),
\end{equation}
the $Q$ function becomes
\begin{eqnarray}
Q(\psi, \psi^m) = 
-\frac{1}{2}  \sum_{j=1}^N \int_{\bs{z}^j \vert \bs{y}^j, \psi^m} \left [ \log \vert \bs{\Sigma} \vert + (\bs{z}^j-\bs{X}^j \bs{\beta})' \bs{\Sigma}_L^{-1} (\bs{z}^j-\bs{X}^j \bs{\beta}) \right ] \cdot \mbox{TMN}(\bs{z}^j, A^j,  \bs{X}^j \bs{\beta^m}, \bs{\Sigma}^m) \de \bs{z}^j.
\label{eq:Qfun3}
\end{eqnarray}
The addends in the square brackets of \eqref{eq:Qfun3} lead to two terms, the first of which can be simplified as
\begin{align}
-\frac{1}{2} \log \vert \bs{\Sigma} \vert \sum_{j=1}^N  \int_{\bs{z}^j \bs{y}^j, \psi^m} \mbox{TMN}(\bs{z}^j, A^j,  \bs{X}^j \bs{\beta^m}, \bs{\Sigma}^m) \de \bs{z}^j 
= -\frac{N}{2} \log \vert \bs{\Sigma} \vert.
\label{eq:Qterm1} 
\end{align}
By the cyclicity property of the trace of a matrix
\begin{equation}
(\bs{z}^j-\bs{X}^j \bs{\beta})^\trans \bs{\Sigma}^{-1} (\bs{z}^j-\bs{X}^j \bs{\beta}) = \Tr \{ \bs{\Sigma}^{-1} (\bs{z}^j-\bs{X}^j \bs{\beta}) (\bs{z}^j-\bs{X}^j \bs{\beta})^\trans \},
\label{eq:quadforms}
\end{equation}
hence the second term of \eqref{eq:Qfun3} can be written as
\begin{align}
\nonumber & -\frac{1}{2} \sum_{j=1}^N \int_{\bs{z}^j \vert \bs{y}^j, \psi^m} \Tr \{ \bs{\Sigma}^{-1} (\bs{z}^j-\bs{X}^j \bs{\beta}) (\bs{z}^j-\bs{X}^j \bs{\beta})^\trans \} \cdot \mbox{TMN}(A^j,  \bs{X}^j \bs{\beta^m}, \bs{\Sigma}^m) \de \bs{z}^j \\
\nonumber &= -\frac{1}{2} \Tr  \left \{ \bs{\Sigma}^{-1} \sum_{j=1}^N \int_{\bs{z}^j \vert \bs{y}^j, \psi^m} (\bs{z}^j-\bs{X}^j \bs{\beta}) (\bs{z}^j-\bs{X}^j \bs{\beta})^\trans \cdot \mbox{TMN}(A^j,  \bs{X}^j \bs{\beta^m}, \bs{\Sigma}^m) \de \bs{z}^j  \right \} \\
&\equiv -\frac{1}{2} \Tr \left \{ \bs{\Sigma}^{-1} \sum_{j=1}^N \mathbb{E}_{\bs Z^j \vert \bs Y^j, \psi^m} \left[ (\bs{Z}^j-\bs{X}^j \bs{\beta}) (\bs{Z}^j-\bs{X}^j \bs{\beta})^\trans \right]  \right \}.
\label{eq:Qterm2}
\end{align}
By combining equations \eqref{eq:Qterm1} and \eqref{eq:Qterm2} we obtain the final expression for the $Q$ function as in equation \eqref{eq:qem}
\begin{eqnarray}
\nonumber Q(\psi, \psi^m) &=& -\frac{N}{2} \left [ \log \vert \bs \Sigma \vert
+ \Tr  \left \{ \bs \Sigma^{-1} \frac{1}{N} \sum_{j=1}^N { \mathbb{E}_{\bs{Z}^j \vert \bs{Y}^j, \psi^m} \left \{ (\bs{Z}^j - \bs{X}^j \bs{\beta} )(\bs{Z}^j - \bs{X}^j \bs{\beta} )^\trans  \right \} } \right \} \right ].
\label{eq:appQfun}
\end{eqnarray}
%
%%%%%%%%%%%%%%%%%%%%%%%%%%%%%%%%
\section{Algorithms}
\label{app:algorithms}

\begin{algenv}
{SMC sampler  
	\label{alg:SMCsampler}
	}
{Key steps of a SMC sampler with a Random Walk Metropolis transition kernel and normalising constant estimation \vspace{2ex}}
Initialisation: & \twocolpf{obtain a weighted particle approximation from the initial distribution}
{ (W_0^{(k)}, \bs{Z}_0^{(k)}) \thicksim \pi_0(\theta_0) 
\label{algSMC:a} 
} 
\\
& \twocolp{set parameters $\theta_1, \bs{\Sigma}^{\rm MH}_1, \kappa_1, \zeta_1, \xi_1$ for a first move, $n=1$}
\\
SMC core: & \twocolp{Repeat the following loop until $\theta_n \equiv \theta_T \ (\pi_n \equiv \pi_T)$} 
\\
Loop: 
& \twocolpf{evaluate incremental weights}
{\tilde{w}_n(\bs{Z}_{n-1}^{(k)},\bs{Z}_{n-1}^{(k)}) = \frac{\gamma_n{(\bs{Z}_{n-1}^{(k)})}}{\gamma_{n-1}(\bs{Z}_{n-1}^{(k)})} 
\label{algSMC:c} } 
\\
& \twocolpf{update normalised weights}
{W_n^{(k)} \propto W_{n-1}^{(k)} \tilde{w}_n^{(k)} 
\label{algSMC:d} 
} 
\\
& \twocolpf{update normalising constant estimate}
{\widehat{C}_n = \widehat{C}_{n-1} \sum_{k=1}^M {W_{n-1}^{(k)} \tilde{w}_n^{(k)} } 
\label{algSMC:e} 
} 
\\
& \twocolpf{evaluate $\mbox{ESS}_n$ as a measure of the degree of degeneracy}
{\mbox{ESS}_n = \frac{1}{\sum_{k=1}^M \limits {(W_n^{(k)})^2 }} 
\label{algSMC:f} 
} 
\\
& \twocolpf{if ESS $<$ ESS$^*$ resample}
{(W_n^{(k)}, \bs{Z}_{n-1}^{(k)}) \to \left(\frac{1}{M}, \tilde{\bs{Z}}_{n-1}^{(k)} \right) \thicksim \pi_n 
\label{algSMC:g} 
} 
\\
& \twocolp{MCMC step: $\forall k \in \{ 1, 2, \ldots, M \}$} \\
& \twocolp{\hspace{5em} sample $\bs{Y}^k \thicksim \mathcal{N}( \bs{Z}_{n-1}^{(k)}, \bs{\Sigma}^{\rm MH}_n)$} \\
& \twocolpf{\hspace{5em} set $\bs{Z}_n^{(k)} = \bs{Y}^k$ with probability}{\alpha^k = 1 \wedge \rho^k 
\label{algSMC:h}
} 
\\
& \twocolpf{\hspace{5em} where} {\rho^k = \frac{\pi_n(\bs{Y}^k)} {\pi_n(\bs{Z}_{n-1}^{(k)} )} 
\equiv \frac{\gamma_n(\bs{Y}^k)} {\gamma_n(\bs{Z}_{n-1}^{(k)} )} 
\label{algSMC:i} 
} 
\\
& \twocolpf{adapt scaling factor}{
\log(\kappa_{n+1}) = \log(\kappa_{n}) + \xi_n (\hat{\alpha}_n(\log(\kappa_{n})) - \alpha^{\star })
\label{eq:sfstochapp}
}
\\
& \twocolp{set new proposal covariance matrix $\bs{\Sigma}^{\rm MH}_{n+1} = \kappa_n \widehat{\bs{\Sigma}}_{\pi_{n}}$}
\\
& \twocolpf{update the parameter identifying the next target}{\theta_{n+1} = \left[
		\theta_{n} + 
			\left( 
				\zeta_n 
 					\frac{\mbox{ESS}_n -
\mbox{ESS}^{\star}_A} {M}
				\vee \Delta \theta_{\min}
			\right) \right]
					\wedge \theta_T ,
\label{algSMC:l}
} \\
& \twocolpf{current particle approximation}
{(W_n^{(k)}, \bs{Z}_n^{(k)}) \mbox{ ``} \thicksim \mbox{'' } \pi_n 
\label{algSMC:j} 
} 
\\
& \twocolpf{go to next iteration}{n=n+1} % \ (n=n+1)
\\
End of loop: & \twocolpf{particle approximation available}
{(W_{T}^{(k)}, \bs{Z}_{T}^{(k)}) \mbox{ ``} \thicksim \mbox{'' } \pi_T 
\label{algSMC:k} 
} 
\\
\end{algenv}

\paragraph{Further implementation details.}
When targeting the multivariate Student $t$ distribution truncated to a domain $A$ as described in Section \ref{sec:smctmn} the parameter $\theta=A$ can be defined as a vector of components $\theta = (\pm a_1, \ldots, \pm a_p)^{\trans}$ with the signs and direction of trunctation given by the observations. Similarly a vector $\theta_n = (\pm a_{1n}, \ldots, \pm a_{pn})^{\trans}$ defines the target region $A_n$ at iteration $n$. In practice the algorithm cycles through the dimensions one at the time, so we can focus on one particular component for a more detailed description. Drawing from an untruncated distribution effectively means to fix $a_{i0}=-\infty$. The first truncation points $a_{i1}$ can then be chosen for example by ensuring that a certain proportion of the probability mass of a multivariate normal distribution with independent components is preserved after the truncation (to make sure that a non-negligible number of particles is kept).  After the initialization the algorithm proceeds by updating each component according to equation \eqref{algSMC:l} in Algorithm \ref{alg:SMCsampler}. The initial covariance matrix for the random walk Metropolis $\bs{\Sigma}^{\rm MH}_1$ is set equal to the covariance matrix target of the multivariate normal distribution (untruncated).  Further tuning parameters in our runs were set as $\kappa_1=1, \zeta_1=2 , \xi_1=7 $, $\Delta \theta_{\min}=.02$, however they will depend on the particular scale of the problem, but they are not automatically tuned in our implementation. The resampling and adaptive thresholds ESS$^*$ and  ESS$^{\star}_A$ are both set to $.8M$.

\begin{algenv}
{Probit SMC EM  
	\label{alg:SMCEM}
	}
{Key steps of the SMC EM algorithm for multivariate probit models \vspace{2ex}}
Initialisation: & \twocolpf{Set parameters}
{ \bs{\psi}^0 = (\bs{\Sigma}^0, \bs{\beta}^0), \quad m=0  
\label{algSMCEM:a} 
} 
\\
 & \twocolpf{obtain a sample (possibly weighted) from the initial distribution}
{ (W_0^{(k)}, \bs{Z}_0^{(k)}) \thicksim \pi_0(\bs{\psi}^0),
}
\\
EM core: & \twocolp{Repeat the following loop until \mbox{$\lVert \bs{\psi}^m - \bs{\psi}^{m-1} \rVert$} converges up to noise} 
\\
Loop: & \twocolp{$m=m+1, \tilde{\bs{\beta}}^0 = \bs{\beta}^0, \tilde{\bs{\Sigma}}^0 = \bs{\Sigma}^0, n=0$}\\
M-step: & \twocolp{Cycle through conditional maximisation until \mbox{$\lVert \tilde{\bs{\beta}}^{n} - \tilde{\bs{\beta}}^{n-1} \rVert  < \epsilon$}}
\\
& \hspace{5em} $n=n+1$\\
& \twocolpf{\hspace{5em} update covariance matrix}{
\tilde{\bs{\Sigma}}^{n} = \frac{1}{N} \sum_{j=1}^N { \sum_{k=1}^M { W^{j (k)}  
(\bs{Z}^{j (k)} - \bs{X}^j \tilde{\bs{\beta}}^{n-1} )
(\bs{Z}^{j (k)} - \bs{X}^j \tilde{\bs{\beta}}^{n-1} )^{\trans} } } 
} \\
& \twocolpf{\hspace{5em} update regression coefficients}{\tilde{\bs{\beta}}^{n} = 
\bigg ( \sum_{j=1}^N { (\bs{X}^j)^{\trans} (\tilde{\bs{\Sigma}}^{n})^{-1}  \bs{X}^j }  \bigg )^{-1} 
\sum_{j=1}^N { (\bs{X}^j)^{\trans} (\tilde{\bs{\Sigma}}^{n})^{-1} 
\sum \limits _{k=1}^M {  \left ( W^{j (k)} \bs{Z}^{j (k)} \right ) }  }}
\\ 
& \twocolpf{update parameter $\bs{\psi}_m$ before E-step}{\bs{\Sigma}^m = \tilde{\bs{\Sigma}}^n, \bs{\beta}^m = \tilde{\bs{\beta}}^n}
\\
E-step: & \twocolp{Implement a SMC sampler to move samples from $\pi_{m-1}(\bs{\psi}^{m-1})$ to target $\pi_{m}(\bs{\psi}^{m})$}
\\
& \twocolp{$\forall j \in \{ 1, 2, \ldots, N \}$} \\
& \twocolp{Rescale sample: $\forall k \in \{ 1, 2, \ldots, M \}$ }
\\
& \twocolpf{}{\tilde{Z}^{(k)}_{m-1} = \bs{D}^{-1} Z^{(k)}_{m-1}} \\
& \twocolpf{\hspace{5em} with scaling $\bs{D}$ such that}{\bs{X}^j \beta^m = \bs{D}^{-1} \bs{X}^j \beta^{m-1}} \\
& \twocolpf{\hspace{5em} set covariance matrix}{\tilde{\bs{\Sigma}}^{m-1} = \bs{D}^{-1} \bs{\Sigma}^{m-1} \bs{D}^{-1}} \\
& \twocolpf{\hspace{5em} current particle approximation}
{(W_{m-1}^{(k)}, \tilde{\bs{Z}}_{m-1}^{(k)}) \mbox{ ``} \thicksim \mbox{'' } \mbox{TMN}(A^j, \bs{X}^j \bs{\beta}^m, \tilde{\bs{\Sigma}}^{m-1})
\label{algSMCEM:b}
}\\
& \twocolpf{build a SMC sampler to move from}{ \pi_0 = \mbox{TMN}(A^j, \bs{X}^j \bs{\beta}^m, \tilde{\bs{\Sigma}}^{m-1}), \quad \theta_0 = \tilde{\bs{\Sigma}}^{m-1} } \\
& \twocolpf{\hspace{5em} to}{ \pi_T = \mbox{TMN}(A^j, \bs{X}^j \bs{\beta}^m, \bs{\Sigma}^{m}), \quad \theta_T = \bs{\Sigma}^{m} } 
\\
\end{algenv}

\section{Log-likelihood correction} \label{logcorrection}

The log-likelihood is the sum of the log-probabilities of the regions corresponding to each observation, for which the SMC sampler provides noisy estimates. Assume the value returned is $p_j(1+\xi_j)$ for each observation $j$, where $\xi_j$ are random relative noise variables with zero mean so that any bias is kept in the value $p_j$.  The total log-likelihood is then
\begin{equation}
l = \sum_j \log\left[p_j(1+\xi_j)\right] = \sum_j \log(p_j) + \log(1+\xi_j) = \sum_j \log(p_j) + \xi_j -\frac{\xi_j^2}{2}+
\ldots
\end{equation}
with the logarithms expanded up to second order.  Consider the sums over the random noises and their squares.  From the central limit theorem, these should tend towards normal distributions with variances related to the sums of the second and fourth moments of the $\xi_j$. Assuming the relative errors $\xi_j\ll 1$, fourth and higher moments will decay quickly compared to the second, leading to the approximation
\begin{equation}
\sum_j \log(1+\xi_j) \approx \zeta - \frac{\sigma^2}{2},
\end{equation}
where $\sigma^2$ corresponds to the variance between different runs of the log-likelihood estimate and $\zeta$ is a normally distributed random variable with the same variance $\zeta\sim N(0,\sigma^2)$.  The log-likelihood should then be corrected to
\begin{equation}
\hat{l} = \sum_j \log(p_j) \approx \sum_j \log\left[p_j(1+\xi_j)\right] + \frac{\sigma^2}{2}.
\end{equation}

\bibliographystyle{model2-names}
\bibliography{smcemmpmbib}

\end{document}